\documentclass[aps,pra,reprint,twocolumn,superscriptaddress,longbibliography]{revtex4-2}

\usepackage{graphicx}
\usepackage{xcolor}
\usepackage{dcolumn}
\usepackage{bm}
\usepackage{amsfonts}
\usepackage{amsmath}
\usepackage{amssymb}
\usepackage{amsthm}
\usepackage{natbib}
\usepackage{physics}
\usepackage{comment}
\usepackage{appendix}
\usepackage[normalem]{ulem}
\usepackage{hyperref}
\usepackage{orcidlink}
\hypersetup{
 colorlinks=true,
 linkcolor=blue,
 filecolor=blue,
 citecolor=blue, 
 urlcolor=cyan,
}
\usepackage[capitalise]{cleveref}
\usepackage{verbatim}
\usepackage[strings]{underscore}

\def \addCQuIC {Center for Quantum Information and Control, University of New Mexico, Albuquerque, NM, 87131, USA}
\def \addPandAUNM {Department of Physics and Astronomy, University of New Mexico, Albuquerque, NM, 87106, USA}
\def \addStrathclyde {Department of Physics, SUPA and University of Strathclyde, Glasgow G4 0NG, United Kingdom}
\begin{document}
\title{Quantum optimal control of the Dicke manifold in dipolar Rydberg atom arrays}

\author{Ivy Pannier-G\"unther\orcidlink{0009-0003-9970-1658}}
\email{igunther@unm.edu}
\affiliation{\addCQuIC} \affiliation{\addPandAUNM}

\author{Vikas Buchemmavari\orcidlink{0000-0002-1592-5626}}
%\email{bsdvikas@unm.edu}
\affiliation{\addCQuIC} \affiliation{\addPandAUNM}

\author{Pablo M. Poggi\orcidlink{0000-0002-9035-3090}}
\affiliation{\addStrathclyde} \affiliation{\addCQuIC} 

\author{Ivan H. Deutsch\orcidlink{0000-0002-1733-5750}}
\affiliation{\addCQuIC} \affiliation{\addPandAUNM}
\date{2026-05-31}

\begin{abstract}
The ability to engineer and control quantum states of many-body systems is a central challenge in quantum information science. For a register of $N$ qubits, the full Hilbert space dimension grows exponentially as $2^N$, rendering generic state preparation and control infeasible without exploiting structure or symmetry. A particularly important and physically motivated restriction is to the fully symmetric subspace, spanned by the Dicke states, which are simultaneous eigenstates of collective spin $J=N/2$. Ensembles of Rydberg atoms interacting via electric dipoles in two-dimensional tweezer arrays form a promising platform for achieving such control. However, the finite range of dipole-dipole interactions poses a challenge to generating and controlling the Dicke manifold because the Hamiltonian incurs leakage from the computational subspace. To counteract this leakage, we perform quantum optimal control algorithms on a truncated Hilbert space according to our newly developed method of ``irrep distillation'' (IRD), which captures the process by which the symmetric subspace couples to leakage error-spaces, using only linear-scaling Hilbert dimension~\cite{IvyPG25}. We implement gradient ascent pulse engineering (GrAPE) on control schemes with little or no local addressing, to generate resourceful states like Greenberger-Horne-Zeilinger, Dicke, and extremal quantum states. We benchmark each scheme of IRD-GrAPE for its quantum speed limit (QSL), as well as exactly testing pulse fidelities on small system sizes and predicting fidelities using higher-order IRD on larger systems.
\end{abstract}
\maketitle

\section{Introduction}\label{sec:intro}
Quantum control within the permutationally symmetric subspace of $N$ spin-1/2 particles (qubits), also known as the Dicke manifold~\cite{Dicke54, Kirton19}, has a number of important applications in quantum information processing (QIP). Entangled probe states in the symmetric subspace, such as spin squeezed states~\cite{Sinatra22,Pezze18}, Dicke states~\cite{Saleem24,Zhang14}, and extremal quantum states~\cite{Goldberg18,Goldberg20,Martin2020optimaldetectionof,Goldberg21} are resources for many quantum sensing and metrology protocols~\cite{Marconi26}. In addition, states within the symmetric subspace form a collective spin, $J=N/2$, in which one can encode quantum information in a synthetic qudit of dimension $d=2J+1=N+1$. Qudit-based QIP is a subject of much recent interest for computing~\cite{Wang20,Wang25}, simulation~\cite{Bauer23PRX,Bauer23Nature,blatt2012quantum,meth2025simulating}, and communication~\cite{Esposito23,yu2025quantum,Spencer2026quditlowdensity}. A goal, thus, is to have the capacity to prepare arbitrary states and implement desired unitary gates on the Dicke manifold.

Direct control of the Dicke manifold arises in systems where the Hamiltonian is permutationally symmetric, equivalent to uniform all-to-all couplings between the qubits. This can be achieved, e.g., in cavity QED, where atoms couple uniformly to a common single mode of the electromagnetic field~\cite{TavisCummings} and in an ensemble of atoms symmetrically coupled within a Rydberg blockade radius~\cite{Brion07}. For natural systems, with finite range interactions, such direct control is not possible. However, in systems with long-range interactions falling off as $1/r^\alpha$ arranged in a $D$-dimensional lattice, such as trapped arrays of Rydberg atoms~\cite{labuhn2016tunable,Zeiher17,Hollerith22} and polar molecules~\cite{yan2013observation}, new opportunities arise due to properties of the many-body spin Hamiltonian. These include quantum many-body scars (QMBS) in Ising models~\cite{foss2016entanglement, LP25}, $U(1)$ symmetry and tower states in $XX$ and $XXZ$ models~\cite{Comparin2022PRA, Perlin2020PRL, young2023enhancing}, and finite temperature symmetry breaking in such models~\cite{Block2024NatPhys, Roscilde2024PRL, bornet2023scalable}. In all of these cases, many-body spin systems interacting over ``long" (albeit finite) ranges can exhibit approximately all-to-all connected behavior, with either similar or identical scaling in their collective dynamics~\cite{LP20,Defenu24,Mori19,Zheng_Senoo_Adiabatic_echo_2026}. Hilbert space fragmentation that arises for these models allows for quench dynamics that constrain the system to the Dicke manifold over a sufficiently long timescale, and the natural dynamics generate spin squeezing.

More general control of the Dicke manifold requires time-dependent driving of the system. Quantum optimal control (QOC) provides a set of tools for generating unitaries and states on a chosen Hilbert space by pulse engineering at the Hamiltonian level~\cite{Ansel24,duncan2025}. This method tailors algorithms for state/unitary generation to the operations that are simple and natural to the hardware. Quantum optimal control has been applied to Rydberg atomic spin-squeezing~\cite{Carrera25,li2026optimal}. More general resource states are a current target for Rydberg-based quantum optimal control theory~\cite{Carrera26}, and thusfar only simulated on small spin-chains with nearest-neighbor interactions~\cite{Muratori25}.

QOC on the full Hilbert space of $N$ qubits grows intractable as $N$ grows. For long-range interacting models one can combine effects of many-body ordering such as QMBS and symmetry breaking to restrict dynamics to a tractable subspace. The challenge is that in the presence of driving, Hilbert space fragmentation is imperfect, and leakage into the the exponentially large Hilbert space must be suppressed. In recent work, Li {\em et al.} showed how, with the addition of global transverse field driving to the dipolar ($\alpha=3$) $XX$ model, one can use QOC to achieve spin squeezing that exceeds the two-axis-twisting benchmark~\cite{li2026optimal}. They employed rotor-spin-wave theory~\cite{Roscilde23} to approximate a tractable Hilbert space, but this method restricts the control to Gaussian squeezing. 

In the work presented here, we extend the application of QOC to pursue {\em universal control} on the Dicke manifold for medium- to long-range interactions. To go beyond Gaussian states, we employ a methodology we have developed dubbed ``irreducible representation distillation," to define a set of irreducible subspaces which represent the finite interaction range as a perturbation on a mean-field Hamiltonian~\cite{IvyPG25}. For a weak perturbation, the number of distilled irreps needed to capture population that leaks out of the Dicke manifold over the course of the evolution is small, and we can approximately truncate the Hilbert space to a dimension that only grows linearly with $N$. With numerically tractable QOC we solve for time-dependent pulses by gradient ascent pulse engineering (GrAPE)~\cite{khaneja2005,motzoi2011} to build resourceful states while negating and avoiding leakage. 

As a particular example, we focus on Dicke manifold control for qubits consisting of two Rydberg states of an atom, with resonant dipolar interactions scaling as $1/r^3$, arranged in 2D tight-packed hexagonal lattice. While in such a 2D lattice the dipolar interactions are not sufficiently long range to yield scalable Dicke manifold control as $N$ gets large, for moderate ensembles we will show that we retain the desired properties, such as QMBS. We show competitive fidelity and quantum speed limits for ensembles of up to 19 atoms, including comparisons against the exact system for up to 14, and with more sophisticated models like higher-order leakage couplings and slight local addressing in the control, we demonstrate that these fidelities become tunable.

The article is organized as follows. In \cref{sec:model}, we outline the system of interest, an ensemble of Rydberg-atomic qubits described by power-law decaying $XX$-interactions and a global driving microwave field. The amplitude and phase of the driving field serve as tunable parameters for quantum optimal control, and the finite range of the interaction incurs leakage out of the computation space. We model this leakage in \cref{sec:IRD} using the formalism of irrep distillation~\cite{IvyPG25}, allowing us to solve control algorithms efficiently in a truncated Hilbert space. We lay out the method of quantum optimal control in \cref{sec:control}, to implement gradient ascent pulse engineering in pursuit of specific symmetric target states. Our control algorithms encompass three strategies: control ignoring leakage in \cref{sec:naive}, leakage-aware global control in \cref{sec:global}, and control with a tunable local parameter in \cref{sec:local}. The results of these control algorithms follow in \cref{sec:res}, and in \cref{sec:fid} we compare the fidelities of resulting state-constructions across each strategy, and against current gate-based algorithms. Next in \cref{sec:QSL}, we estimate the quantum speed limits of our targets and strategies to find that leakage protracts the control, but does not change its proportionality with system size. In \cref{sec:disc} we evaluate the performance of the control for example Rydberg systems, including the effects of the finite Rydberg lifetime. Finally, we summarize and suggest future directions in \cref{sec:conc}.

\section{Model}\label{sec:model}
\begin{figure}
    \centering
    \includegraphics[width=\linewidth]{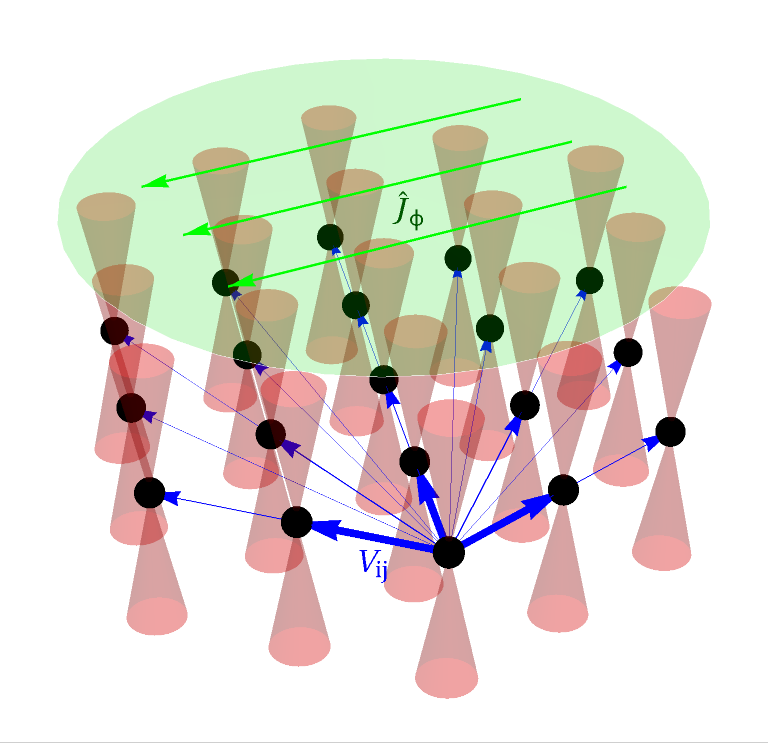}
    \caption{Schematic diagram for quantum optimal control in the Dicke manifold. Rydberg atoms in a 2D array of optical tweezers arranged in a close-packed triangular lattice interact via resonant dipole-dipole interactions with an interaction graph $V_{ij} = \frac{V}{\mathcal{N}}||\mathbf{i}-\mathbf{j}||^{-3}$ (blue). Control is designed by a time-dependent global field which drives Rabi oscillations, generated by the component collective spin, $\hat{J}_\phi$ (green).}
    \label{fig:Coupling}
\end{figure}
The system of interest is an ensemble of pseudospins (qubits) on a regular hexagon-shaped triangular lattice interacting via an $XX$-Hamiltonian with dipolar, $1/r^3$, interactions. For concreteness, this corresponds to neutral atom array assembled in optical tweezers, with resonant electric dipole-dipole interactions for qubits encoded in two opposite-parity Rydberg states. In addition, the ensemble is controlled by a globally uniform resonant microwave field whose intensity and phase determines the Rabi frequency $\Omega(t)$ and $\phi(t)$, which can be modulated in time. The total Hamiltonian takes the form,
\begin{subequations}
\begin{align}
    \hat{H}(t)=&\hat{H}_\mathrm{con}(\phi(t),\Omega(t))+\hat{H}_\mathrm{int},\label{eq:Ht}\\
    \hat{H}_\mathrm{con}(\phi(t),\Omega(t))\equiv & \frac{\Omega(t)}{2}\sum_{i=1}^{N}\left[\cos\phi(t) \hat{\sigma}_x^{(i)}+\sin\phi(t)\hat{\sigma}_y^{(i)}\right]\nonumber\\
    = & \Omega(t)\hat{J}_{\phi(t)},\\
    \hat{H}_\mathrm{int}\equiv &\frac{1}{4}\sum_{i,j=1}^{N}V_{ij}(\hat{\sigma}_x^{(j)}\hat{\sigma}_x^{(j)}+\hat{\sigma}_y^{(i)}\hat{\sigma}_y^{(j)}),
\end{align}
\end{subequations}
wherein $\hat{J}_{\phi(t)} \equiv \cos{\phi(t)}\hat{J}_x+\sin{\phi(t)}\hat{J}_y$ for collective angular momentum $\hat{J}_\mu=\frac{1}{2}\sum_{i=1}^N\hat{\sigma}_\mu^{(i)}$. For a hexagonal lattice there are $N=3L^2+3L+1$ qubits, with $6L$ sites on the perimeter, as shown in \cref{fig:Coupling}. The interactions between sites on this lattice form a non-sparse weighted graph with Kac-normalized weights given by the adjacency matrix $V_{ij}=\frac{V}{\mathcal{N}}||\mathbf{i}-\mathbf{j}||^{-3}$~\cite{Perlin20,Kuriyattil25}, and interaction energy $V_{1,2}=\frac{V}{\mathcal{N}}=d^2/a^3$ for lattice spacing $a$ and dipole matrix element $d$.

\begin{figure}
\centering
\includegraphics[width=\linewidth]{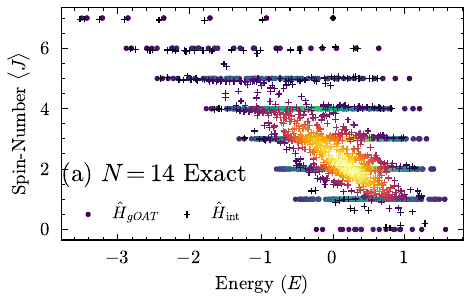}
\includegraphics[width=\linewidth]{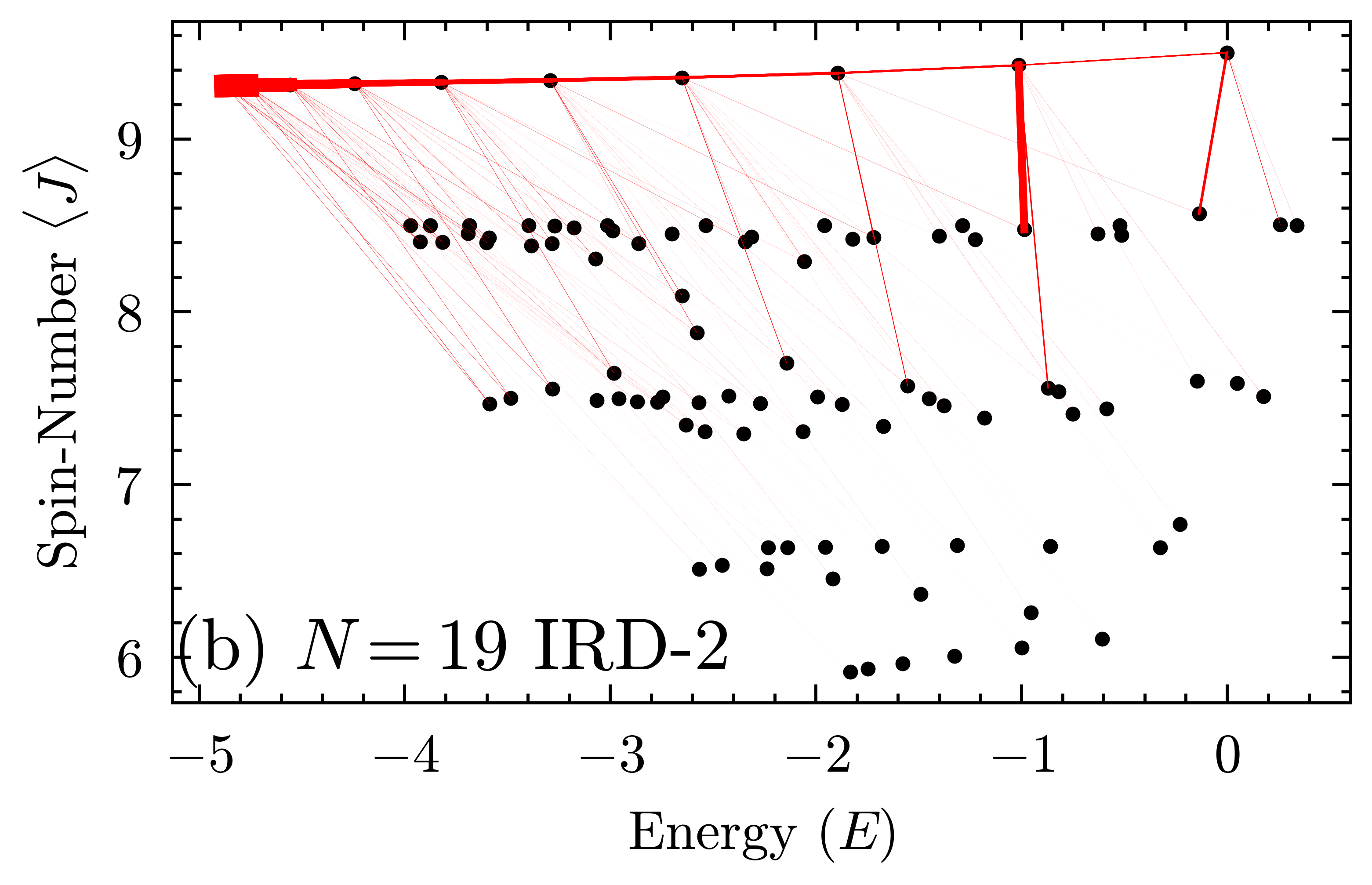}
\caption{\label{fig:DistSpec} (a) Eigenstates of the dipole-dipole interaction Hamiltonians $\hat{H}_\mathrm{int}$ and $\hat{H}_{gOAT}$, on triangular lattices of $N=14$ two-level atoms. Eigenstates are plotted for eigenenergy $E$ and spin-number expectation $\langle J\rangle$ defined such that $\langle J\rangle^2+\langle J\rangle=\langle \mathbf{J}^2\rangle$. In purple-to-green, the gapped OAT, for which $J$ is an exact quantum number. In black-to-orange, the dipole-dipole interaction, which forms a QMBS approximating the Dicke manifold. (b) Eigenstates of $\hat{H}_\mathrm{int}$ on $\mathcal{H}_D^{(2)}$ (using irrep distillation explained in \cref{sec:IRD}) for $N=19$, with red bonds showing QMBS off-diagonal couplings caused by the control Hamiltonian, $\hat{H}_\mathrm{con} = \Omega(t) \hat{J}_{\phi(t)}$.}
\end{figure}

To understand the how long-range interactions can enable collective spin control, it is useful to express the interaction Hamiltonian as,
\begin{subequations}
    \begin{align}
        \hat{H}_\mathrm{int} =& \hat{H}_{gOAT} + \hat{V}_\mathrm{pert},\label{eq:Hint}\\
        \hat{H}_{gOAT} =& \frac{1}{4}\sum_{i,j}V_{ij} \hat{\vec{\sigma}}^{(i)} \cdot \hat{\vec{\sigma}}^ {(j)}-\frac{V}{N}\hat{J}_z^2,\label{eq:HgOAT}\\
        \hat{V}_\mathrm{pert}=& \frac{1}{4}\sum_{ij} (\frac{V}{N}-V_{ij}) \hat\sigma_z^{(i)}\hat\sigma_z^{(j)}.\label{eq:Vpert}
    \end{align}
\end{subequations}
The one-axis twisting (OAT) Hamiltonian, $(V/N) \hat{J}_z^2$, is the canonical interaction used to generate spin squeezing when one has access to all-to-all interactions~\cite{Kitagawa1993}. For finite-range interactions, the ``gapped OAT'' Hamiltonian, $\hat{H}_{gOAT}$, introduced in~\cite{young2023enhancing}, has $U(1)$ symmetry and an energy gap between states with different total spin $J$, thereby enabling spin squeezing without full spin coordination. The robustness of the squeezing dynamics is determined by the energy gap. This is perturbed by $\hat{V}_\mathrm{pert}$, which quantifies the spin coordination number in relation to an all-to-all interaction. For a generic graph $V_{ij}$, the size of the spectral gap $\Delta$ is determined by the Laplacian matrix $\Lambda_{ij}=\bar{V}_i\delta_{ij}-V_{ij}$, where $\bar{V}_i=\sum_j V_{ij}$ \cite{Kuriyattil25}. In the limit $N\rightarrow \infty$, $\Delta$ remains large for power-law interactions $1/r^\alpha$ only for lattices in dimension $D\geq\alpha$. Nonetheless, for finite $N$, the gap remains substantial even for $N=19$, $\alpha=3$, $D=2$, as discussed in \cref{app:Lap}. The spectrum of $\hat{H}_\mathrm{int}$ is shown in \cref{fig:DistSpec}(a), displaying these features.

In the limit of all-to-all interactions, the Hamiltonian $\hat{H}_0(t) =\Omega(t) \hat{J}_{\phi(t)}-(V/N)\hat{J}_z^2$ enables universal control of a spin $J$ on a $2J+1$-dimensional irrep, as demonstrated in \cite{Omanakuttan21,Chaudhury07}. The Hamiltonian $\hat{H}(t)$ in \cref{eq:Ht} approximates $\hat{H}_0(t)$ when restricted to the Dicke manifold, thanks to a set of QMBS with spin $\langle J\rangle\approx N/2$~\cite{LP25}. To the degree that $\langle J\rangle$ is not an exact quantum number, the control term $\Omega(t) \hat{J}_{\phi(t)}$ will couple states out of the symmetric subspace. These off-diagonal matrix elements are shown in \cref{fig:DistSpec}(b) as couplings between QMBS states and other eigenstates. The result is leakage out of the Dicke manifold, potentially to the exponentially large Hilbert space. 

The system size $N=19$ is still too large for exact diagonalization, and the triangular lattice renders matrix product state (MPS) methods inefficient, opening the problem to new formal methods. Tensor networks with belief propagation~\cite{Tindall24,Tindall25,Midha26} are making rapid strides on multidimensional dynamics, but still rely on long loops and short-range correlations in the lattice. Outside tensor networks, approaches include time-dependent spin-waves (TDSW) and their refinement in the rotor-magnon model of many-spin systems~\cite{Roscilde23,LP25}. But TDSW assumes a strongly polarized state, whose net polarization forms the origin of a bosonic phase space under a Holstein-Primakoff approximation~\cite{TDSW}. Rotor spin-wave theory has recently shown promising results in QOC on two-dimensional Rydberg atom arrays for scalable spin-squeezing~\cite{li2026optimal}, but the question remains whether such approximations can target non-polarized states beyond spin-squeezing. In the next section, we submit our formalism based on ``irrep distillation" which will allow us to perform quantum control on a tractable subspace of the full Hilbert space.

\subsection{Irreducible Representation Distillation}\label{sec:IRD}%Might later need to do some high-level editing of Sec.IRD
The total Hilbert space of $N$ qubits may be decomposed as,
\begin{equation}
    \mathcal{H}=\mathcal{H}_{1/2}^{\otimes N}=\bigoplus_{J=0,1/2}^{N/2}\mathcal{H}_{J}\otimes \mathcal{C}_{N/2-J},\\
\end{equation}
where $\mathcal{H}_J$ is a generic irrep of $SU(2)$ for spin $J$. With the sole exception of $J={N/2}$, these irreps are degenerate, and we track this degeneracy by use of auxiliary vector spaces $\mathcal{C}_x$ with dimensions $||\mathcal{C}_x||=\binom{N}{x}-\binom{N}{x-1}$. The case $\mathcal{H}_{J=N/2}$ is the Dicke manifold (symmetric subspace), spanned by the $N+1$ Dicke states,
\begin{equation}
\ket{J=\frac{N}{2},M} \equiv \ket{D^{(N)}_{k=J-M}}\equiv\{\ket{\uparrow_z}^{\otimes N-k} \ket{\downarrow_z}^{\otimes k} \}_\mathrm{sym},\label{eq:Dicke}
\end{equation}
where the final form is a symmetric superposition of all permutations of the computational basis states. 

Vectors in the degeneracy-tracking degrees of freedom $\vec{u}\in \mathcal{C}_x$ determine under which specific qubit-permutations a state is asymmetric. Thus, there exists a bijection between qubit basis states $\ket{\vec{s}}=\ket{s_1,...s_N}$ for $s\in\{\uparrow,\downarrow\}$ and degenerate irrep states $\ket{J,M,u}=\ket{J,M}\otimes \vec{u}$. Typically this bijection is given by the Clebsch-Gordan series for the addition of angular momentum, which can \textit{efficiently} construct some but not all collective states. However, the Clebsch-Gordan serial basis of irreps is not natural for $\hat{H}_\mathrm{int}$, which is strongly non-diagonal in that basis of each $\mathcal{C}_x$. On the other hand, $\hat{H}_{gOAT}$ in \cref{eq:HgOAT} is diagonalizable in the irrep basis, such that $J$ is an exact quantum number of its eigenstates.

The method of irreducible representation distillation (IRD) efficiently defines a truncated Hilbert space from perturbative principles to construct the irreps based on the Hamiltonian dynamics at hand~\cite{IvyPG25}. We treat $\hat{H}_\mathrm{int}$ as a perturbed form of $\hat{H}_{gOAT}$, given in \cref{eq:HgOAT}, and solve the vectors $\vec{u}\in\mathcal{C}_x$ to maximize and isolate the matrix elements $\bra{\frac{N}{2},M}\hat{H}_\mathrm{int}\ket{\frac{N}{2}-x,M,u}=\bra{\frac{N}{2},M}\hat{V}_\mathrm{pert}\ket{\frac{N}{2}-x,M,u}$. Since the symmetric subspace includes both the initial state and all target states, and since the Hamiltonian is chaotic (which implies that population exiting the symmetric subspace dynamically is unlikely to return in subexponential time), we may treat the perturbation as incurring leakage errors, which IRD models at low order.

Whereas Pauli operators and Pauli strings form the natural operator basis of $\mathcal{H}_{1/2}^{\otimes N}$, we use an operator basis of generalized irreducible spherical tensor operators when considering the irrep picture. These spherical tensors are defined as $\hat{T}^{(k)}_q [J,u;J',v]\equiv$
\begin{equation}
    \sqrt{\frac{2k+1}{2J+1}}\sum_{M=-J'}^{J'} C^{J,M+q}_{J',M;k,q}\ket{J,M+q,u}\bra{J',M,v}
\end{equation}
with matrix elements from Clebsch-Gordan coefficients $C^{J,M}_{j,m;j',m'}$ in accordance with the Wigner-Eckart theorem. These spherical tensors are orthogonal with each other, they have unit Hilbert-Schmidt norm, they are covariant with other $k$-rank tensors under $SU(2)$, and they map between single irreps $\hat{T}^{(k)}_q:\mathcal{H}_{J'}\otimes \vec{v}\to \mathcal{H}_{J}\otimes \vec{u}$ for $\vec{u}\in\mathcal{C}_{N/2-J}$ and $\vec{v}\in\mathcal{C}_{N/2-J'}$. All these desirable qualities make them an attractive operator basis for the present work.

All three of $\hat{H}_\mathrm{int}$, $\hat{H}_{gOAT}$, and $\hat{V}_\mathrm{pert}$ in \cref{eq:Hint,eq:HgOAT,eq:Vpert} are spanned by tensors of the form $\hat{T}^{(0)}_0$ and $\hat{T}^{(2)}_0$, and due to Clebsch-Gordan constraints this implies that $\bra{J',M',v}\hat{H}_\mathrm{int}\ket{J,M,u}\neq 0$ only if $|J-J'|\leq 2$. We define matrix weights $F_0$ and $F_2$ as the inner product of the interaction with spherical tensors of these two ranks,
\begin{align}
    F_0(x,u,v)\equiv&\Tr(\hat{H}_\mathrm{int}\hat{T}^{(0)}_0[\frac{N}{2}-x,u;\frac{N}{2}-x,v]),\\
    F_2(x,u;y,v)\equiv&\Tr(\hat{H}_\mathrm{int}\hat{T}^{(2)}_0[\frac{N}{2}-x,u;\frac{N}{2}-y,v]),
\end{align}
which we may treat as linear maps between degeneracy spaces, $\hat{F}_0:\mathcal{C}_x\to \mathcal{C}_x$ and $\hat{F}_2:\mathcal{C}_x\to\mathcal{C}_y$. Ergo, first-order perturbation theory only requires solutions to
\begin{align}
    &\bra{\frac{N}{2},M}\hat{H}_\mathrm{int}\ket{\frac{N}{2},M}\nonumber\\
    =&F_0(0)\bra{\frac{N}{2},M}\hat{T}^{(0)}_0\ket{\frac{N}{2},M}\nonumber\\
    +&F_2(0;0)\bra{\frac{N}{2},M}\hat{T}^{(2)}_0\ket{\frac{N}{2},M},\\
    &\bra{\frac{N}{2},M}\hat{H}_\mathrm{int}\ket{\frac{N}{2}-1,M,u}\nonumber\\
    =&F_2(0;1,u)\bra{\frac{N}{2},M}\hat{T}^{(2)}_0\ket{\frac{N}{2}-1,M,u},\\
    &\bra{\frac{N}{2},M}\hat{H}_\mathrm{int}\ket{\frac{N}{2}-2,M,u}\nonumber\\
    =&F_2(0;2,u)\bra{\frac{N}{2},M}\hat{T}^{(2)}_0\ket{\frac{N}{2}-2,M,u},
\end{align}
in order to perturb eigenstates of $\hat{H}_{gOAT}$ in $\mathcal{H}_{N/2}$ into approximate eigenstates of $\hat{H}_\mathrm{int}$.

Note, $F_2(0;1,u)$ and $F_2(0;2,u)$ are linear mappings $\hat{F}_2:\mathcal{C}_0\to\mathcal{C}_1$ and $\hat{F}_2:\mathcal{C}_0\to \mathcal{C}_2$, both from a 1-dimensional trivial space $\mathcal{C}_0$. So not only can we express these as rectangular matrices, but in fact as vectors $\vec{F}_2(0;1)\in\mathcal{C}_1$ and $\vec{F}_2(0;2)\in\mathcal{C}_2$. First-order IRD solves these vectors and renormalizes them as basis vectors $\vec{F}_2(0;1)\propto \vec{c}^{(0,1)}\in\mathcal{C}_1$ and $\vec{F}_2(0;2)\propto \vec{c}^{(0,2)}\in\mathcal{C}_2$. Then, IRD truncates the Hilbert space $\mathcal{H}$ down to $\mathcal{H}^{(1)}_D\subseteq\mathcal{H}$,
\begin{align}
\mathcal{H}^{(1)}_D=&\mathcal{H}_{N/2}\oplus \mathcal{H}_{N/2-1,0}\oplus \mathcal{H}_{N/2-2,0}\\
    =&\bigoplus_{x=0}^2 \mathcal{H}_{N/2-x,0},
\end{align}
for $\mathcal{H}_{N/2-x,0}\equiv \mathcal{H}_{N/2-x}\otimes \vec{c}^{(0,x)}$. The dimension of this truncated Hilbert space is linear with the system size: $||\mathcal{H}_D^{(1)}||=3(N-1)$, making it an efficient representation of dynamics that remain close to $\mathcal{H}_{N/2}$.

Higher orders of IRD proceed in a similar fashion, coupling via $\hat{F}_2$ and $\hat{F}_0$. Since $\hat{T}^{(2)}_0$ is chiefly responsible for mixing between irreps, we consider first the coupling between $\vec{c}^{(0,1)}$ and $\vec{c}^{(0,2)}$ and further irreps, mediated by the (rectangular) matrices $\hat{F}_2(1;x)$ and $\hat{F}_2(2;x)$:
\begin{align}
    \vec{b}^{(1,x)}=\hat{F}_2(1;x)\cdot\vec{c}^{(0,1)},\\
    \vec{b}^{(2,x)}=\hat{F}_2(2;x)\cdot\vec{c}^{(0,2)}.
\end{align}
These vectors produce $\vec{c}^{(1,x)}\propto \vec{b}^{(1,x)}$ and $\vec{c}^{(2,x)}\propto \vec{b}^{(2,x)}$ after Gram-Schmidt orthonormalization. In similar fashion, the matrices $\hat{F}_0(1)$ and $\hat{F}_0(2)$ are square, and from them we solve
\begin{align}
    \vec{b}^{(3,1)}=\hat{F}_0(1)\cdot\vec{c}^{(0,1)},\\
    \vec{b}^{(3,2)}=\hat{F}_0(2)\cdot\vec{c}^{(0,2)},
\end{align}
before orthonormalizing again. Thus we acquire $(\vec{c}^{(0,1)}, \vec{c}^{(1,1)},\vec{c}^{(2,1)},\vec{c}^{(3,1)})\in\mathcal{C}_1$, $(\vec{c}^{(0,2)}, \vec{c}^{(1,2)},\vec{c}^{(2,2)},\vec{c}^{(3,2)})\in\mathcal{C}_2$, $(\vec{c}^{(1,3)},\vec{c}^{(2,3)})\in\mathcal{C}_3$, and $\vec{c}^{(2,4)}\in\mathcal{C}_4$, along with the trivial symmetric irrep vector $\vec{c}^{(0,0)}=1$. These are all the irreps accessible from the symmetric subspace by 2nd-order perturbative couplings via $\hat{T}^{(2)}_0$ and $\hat{T}^{(0)}_0$. From this it follows that $\mathcal{H}_D^{(2)}\subseteq \mathcal{H}$ takes the form,
\begin{align}
    \mathcal{H}^{(2)}_D=&\mathcal{H}_{N/2}\nonumber\\
    \oplus&(\bigoplus_{u=0}^3\mathcal{H}_{N/2-1,u})\oplus (\bigoplus_{u=0}^3\mathcal{H}_{N/2-2,u})\nonumber\\
    \oplus&(\bigoplus_{u=1}^2\mathcal{H}_{N/2-3,u})\oplus \mathcal{H}_{N/2-4,2}
\end{align}
with dimension $||\mathcal{H}^{(2)}_D||=12N-32$, still linear in $N$ albeit with a larger prefactor. When including coupled irreps from all tensor terms, IRD is a controlled approximation, a trait it inherits from perturbation theory.

Under 2nd-order IRD, we draw a three-fold distinction between the target subspace, which is the symmetric subspace $\mathcal{H}_{N/2}$, the auxiliary subspace $\mathcal{H}_{N/2-1,0}\oplus \mathcal{H}_{N/2-2,0}$, and the leakage subspace which includes all other irreps in $\mathcal{H}_D^{(2)}$. The target and auxiliary subspaces together make up $\mathcal{H}_D^{(1)}$, and population in these subspaces will remain in $\mathcal{H}_D^{(2)}$ when acted upon by $\hat{H}(\phi,\Omega)$ because for all $\ket{\psi}\in \mathcal{H}_{N/2}$, $(\hat{H}(\phi,\Omega))^x\ket{\psi}\in\mathcal{H}_D^{(x)}$. For this reason, they are safe for control simulations, since we can be assured of the accuracy of dynamics within these subspaces. But the escape subspace is coupled outside of $\mathcal{H}_D^{(2)}$ by $\hat{H}(\phi,\Omega)$, and thus physical dynamics on these populations may cause leakage out of our model, so we treat the escape subspace as forbidden to computation, an error zone which we model but do not use in control.

When considering only 1st-order IRD, the above three-fold partition of $\mathcal{H}_D^{(1)}$ collapses to a two-fold distinction, because there is no auxiliary subspace beyond the target subspace of $\mathcal{H}_{N/2}$. In that case, only the target subspace itself is safe for control, and all other irreps fall into the leakage subspace. Specifically, the leakage subspace of $\mathcal{H}_D^{(1)}$ is the auxiliary subspace of $\mathcal{H}_D^{(2)}$.

We may define projectors that separate irreps according to the hierarchy of IRD:
\begin{align}
    \hat{\Pi}^{(0)}&=\sum_M\ket{\frac{N}{2},M}\bra{\frac{N}{2},M}\\&=\mathbb{I}_{N/2},\\
    \hat{\Pi}^{(1)}&=\sum_M\ket{\frac{N}{2}-1,M,0}\bra{\frac{N}{2}-1,M,0}\\&+\ket{\frac{N}{2}-2,M,0}\bra{\frac{N}{2}-2,M,0}\nonumber\\&=\mathbb{I}_D^{(1)}-\hat{\Pi}^{(0)},\\
    \hat{\Pi}^{(2)}&=\mathbb{I}_D^{(2)}-\hat{\Pi}^{(1)}-\hat{\Pi}^{(0)},
\end{align}
wherein $\hat{\Pi}^{(0)}$ projects onto the target subspace, $\hat{\Pi}^{(1)}$ projects onto the escape subspace of $\mathcal{H}_D^{(1)}$ or auxiliary subspace of $\mathcal{H}_D^{(2)}$, and $\hat{\Pi}^{(2)}$ projects onto the escape subspace of $\mathcal{H}_D^{(2)}$.

\section{Quantum Control Protocol}\label{sec:control}

Prior to considerations of optimal control strategies on a quantum system, is the question of that system's controllability itself. Consider a given set of Hamiltonians $\mathbb{H}=\{\hat{H}(\vec{f}(t))\}$ parametrized by time-dependent waveforms $\vec{f}(t)$. For a given total control time $T$, the set of unitaries generated for these Hamiltonians is $\mathbb{U}=\mathcal{T}e^{i\int_0^T\mathbb{H}dt}=\{U(\vec f(t),T)\}$, where $\mathcal{T}$ is the time-ordering operator, and the set of states generated from a fiducial initialization is $\mathbf{\Psi}=\mathbb{U}\ket{\psi(0)}$. If $\mathbb{U}$ covers the target space of unitaries, $\mathbb{U}\supseteq \mathcal{U}(\mathcal{H}_\mathrm{tar})$, then the target space is \textit{unitary controllable} by $\mathbb{H}$. This means that $\forall U_\mathrm{tar}\in \mathcal{U}(\mathcal{H}_\mathrm{tar})$, $\exists\vec{f}(t)$ such that for $T>T_*$, $||U_\mathrm{tar}-U(\vec f(t),T)||\le \epsilon,$ $\forall \epsilon $. The time $T_*$ defines the ``quantum speed limit"~\cite{Wiedmann26}. By a looser condition, if $\mathbf{\Psi}$ covers the target Hilbert space $\mathbf{\Psi}\supseteq \mathcal{H}_\mathrm{tar}$, then the target space is \textit{state controllable} by $\mathbb{H}$. Proving either condition is typically difficult, relying on a schematic Lie algebra underlying $\hat{H}(\vec{f}(t))$ with a simple representation on $\mathcal{H}_\mathrm{tar}$.

For all-to-all interactions, the Hamiltonian $\hat{H}_0(\phi(t),\Omega(t))$, with $\mathcal{H}_\mathrm{tar}=\mathcal{H}_{N/2}$, is system is unitarily controllable, as the set $\{\hat J_x,\hat J_y,\hat J_z^2\}$ generates the Lie algebra $SU(2J+1)$~\cite{Chaudhury07}. But for the local case, where $\mathcal{H}_\mathrm{tar}$ is a subspace of the Hilbert space generated by $\hat{H}(\vec{f})$, the nature of the control problem is more complicated. The Lie algebra generated by the non-commuting terms in the Hamiltonian and the topology of resulting group, $\mathbb{U}$, is not straightforward. Topological connectedness of $\mathbb{U}$ strongly influences controllability, and in systems like ours, this remains an open problem~\cite{Carrera26}.  Nonetheless, to the degree that dynamics are restricted to a few irreps, the system will be approximately controllable on the Dicke manifold. Our analysis below demonstrates the efficacy of quantum optimal control for increasingly complex resource states.

\subsection{Optimal Control Algorithm}
We consider here state controllability in the Dicke manifold starting from a fiducial spin coherent state, $\ket{\psi_0}=\ket{\uparrow_z}^{\otimes N}=\ket{\Uparrow_z}=\ket{J=\frac{N}{2},M=\frac{N}{2}}$. Our control algorithms all use gradient ascent pulse engineering (GrAPE), which solves a sequence of time-evolution unitaries parametrized each by the control waveform $(\phi(t),\Omega(t))$ in order to minimize a cost-function $C$ with regard to a target state $\ket{\Psi_\mathrm{tar}}$. For simplicity, we parameterize the waveforms as piecewise constant in time intervals $\delta t$, with a vector of values $\{\phi_k,\Omega_k\}_{k=1}^K$. In each $\delta t =T/K$ we apply a pulse that drives a global rotation of the Bloch vector. For some number of pulses $K$ and total evolution time $T$, is then

\begin{subequations}
\begin{eqnarray}
    \hat{U}=\prod_{k=1}^K \hat{U}_k,\\
    \hat{U}_k=e^{-i\delta t\hat{H}(\phi_k,\Omega_k)},\\
    \ket{\psi_k}=\hat{U}_k\hat{U}_{k-1}...\hat{U}_2\hat{U}_1\ket{\psi_0}.
\end{eqnarray}
\end{subequations}

For a given target state, $\ket{\Psi_\mathrm{tar}}$, we minimize the cost function, defined variously according to our choice of control algorithm. For example, a na\"ive cost function $C_F$ would simply be the infidelity between the final and target states:
\begin{equation}
    C_F=1-F=1-|\braket{\Psi_\mathrm{tar}}{\psi_K}|^2,
\end{equation}
and the minimization always follows the method of gradient descent towards $C\to0$. Gradient descent proceeds by calculating the gradient of the cost function in parameter space $\vec{v}=(\vec{\phi},\vec{\Omega})=(\phi_1,\phi_2,...\phi_{K-1},\phi_K,\Omega_1,\Omega_2,...\Omega_{K-1},\Omega_K)$, such that
\begin{equation}
    \nabla_{\mathcal{V}}C\equiv (\frac{\partial C}{\partial \phi_1},...\frac{\partial C}{\partial \phi_K},\frac{\partial C}{\partial \Omega_1},...\frac{\partial C}{\partial \Omega_K}),
\end{equation}
and beginning from some random initial guess $\vec{v}_0$, iterating the pulse as
\begin{equation}
    \vec{v}_{j+1}=\vec{v}_j-\nabla_{\mathcal{V}}C|_{\vec{v}_j},
\end{equation}
until convergence at a minimum. At each step, calculating $\nabla_\mathcal{V}C$ requires computing each $\frac{\partial \hat{U}_k}{\partial\phi_k}$ and $\frac{\partial \hat{U}_k}{\partial\Omega_k}$, which each in turn entail numerical diagonalizations of $\hat{H}(\phi_k,\Omega_k)$, but which are otherwise analytic. To reduce the possibility of GrAPE settling into a non-global local minimum, we randomly initialize several times and choose the lowest resulting $C$ out of all the runs.

The computational complexity of gradient descent depends strongly on the dimensionality of $\vec{v}\in \mathcal{V}$, with $||\mathcal{V}||=2K$, and thus GrAPE-based control across the full Hilbert space is computationally impractical on large ensembles, where generically the requisite $K$ might scale exponentially with $N$. Moreover, repeated diagonalizations of $\hat{H}(\phi_k,\Omega_k)$, necessary to calculate the gradient, are inaccessible on the full Hilbert space at even modest $N$. Thus, we perform GrAPE in IRD-truncated Hilbert space centered on the computation space $\mathcal{H}_{N/2}$ and its perturbatively coupled neighbors under $\hat{H}_\mathrm{int}$, as this Hilbert space is small enough to remain computationally tractable while still accounting for leakage.

\begin{figure}
    \centering
    \includegraphics[width=0.33\linewidth]{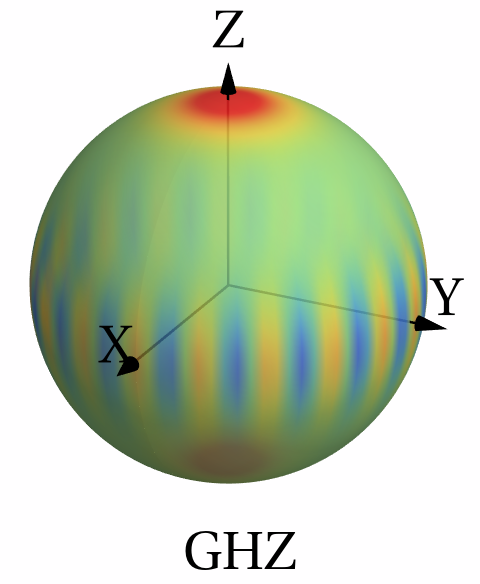}
    \includegraphics[width=0.33\linewidth]{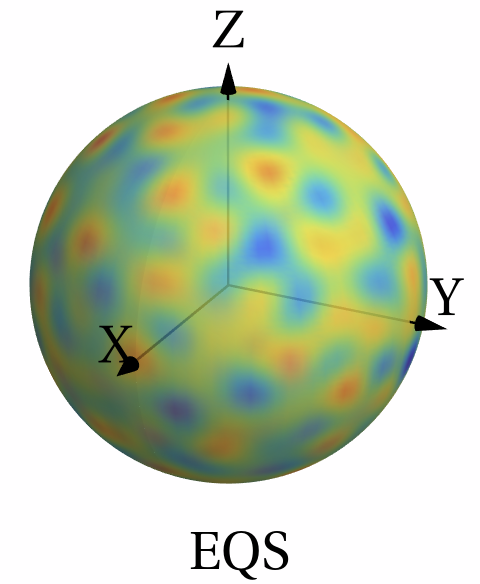}
    \includegraphics[width=0.32\linewidth]{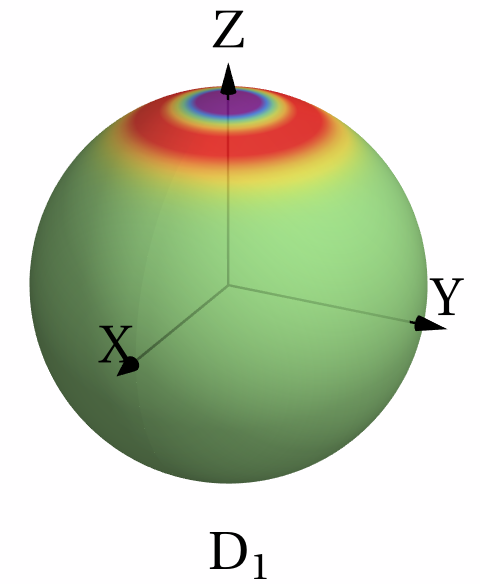}
    \includegraphics[width=0.32\linewidth]{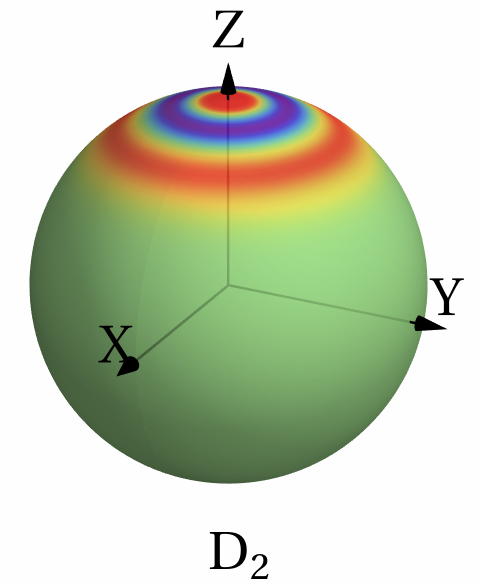}
    \includegraphics[width=0.32\linewidth]{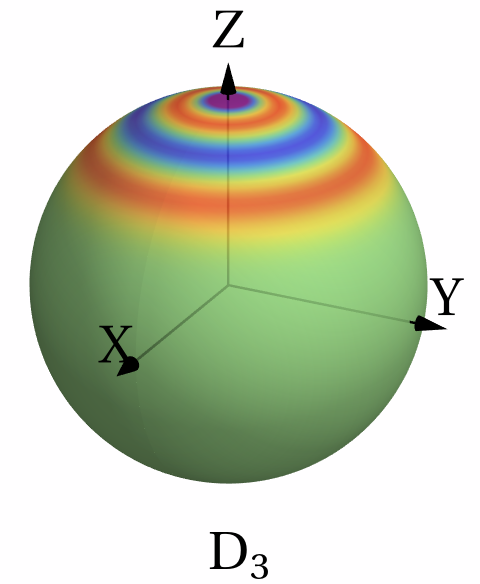}
    \caption{Spin Wigner functions for control target states with $J=\frac{19}{2}$. Top left $\ket{\mathrm{GHZ}}$; top right $\ket{\mathrm{EQS}}$; bottom row $\ket{D_1}$, $\ket{D_2}$, and $\ket{D_3}$.}
    \label{fig:TgtWig}
\end{figure}

We consider the following target states in this work. The simplest targets given $\hat{H}_\mathrm{int}$ are Greenberger-Horne-Zeilinger states $\ket{\mathrm{GHZ}^{(N)}}$,
\begin{equation}
    \ket{\mathrm{GHZ}^{(N)}}=\frac{1}{\sqrt{2}}(\ket{\Uparrow_z}+\ket{\Downarrow_z}),%=\frac{1}{\sqrt{2}}\left(\ket{\frac{N}{2},\frac{N}{2}}+\ket{\frac{N}{2},-\frac{N}{2}}\right),
\end{equation}
known for their maximal sensitivity to $z$-rotations~\cite{Leibfried04,Bollinger96}. Secondly we target Dicke states $\ket{D^{(N)}_k}$ defined in \cref{eq:Dicke}, with increasing complexity as $k$ increases, which can also yield Heisenberg scaling in quantum sensing~\cite{Saleem24}. Finally, for maximum control complexity we consider so-called ``extremal quantum states," conceived of as the ``opposite" to spin-coherent states (which are classically describable), and thus maximally quantum~\cite{Bjork15,Goldberg20}. These are defined in terms of the sum of their low-rank cumulants in the spherical tensor basis,
\begin{equation}
    A^{(N)}_\kappa\equiv \sum_{k=1}^\kappa\sum_{q=-k}^k |\langle T^{(k)}_q[\frac{N}{2};\frac{N}{2}]\rangle|^2,
\end{equation}
such that for a given $N$ and the maximum possible $\kappa$, $\ket{\mathrm{EQS}}$ gives $A^{(N)}_\kappa=0$. The EQS achieves maximum sensitivity to \textit{arbitrary} rotations, although its sensitivity along any specific axis is inferior to that of an appropriate GHZ state~\cite{Bjork15}. Furthermore, EQS are closely related to $t$-designs, and because they have zero expectation value for all low-rank spherical tensors, they mimic that quality of the Haar random ensemble with a single state. We visualize these target states as $SU(2)$ Wigner functions in \cref{fig:TgtWig}.

Whereas GHZ and Dicke states have straightforward vector elements in the irrep basis, solving for the EQS of a given irrep requires sophisticated and computationally intensive symbolic algebra, for which reason we numerically approach the EQS as an optimization problem to minimize the non-negative $A^{(N)}_\kappa$ using gradient descent. We find a class of solutions which give $A^{(19)}_4<10^{-16}$, or zero within machine precision, in agreement with results from ULi\`ege~\cite{LibAC}; the minimum possible result for $\kappa=5$ is $A^{(19)}_5\approx 3\times 10^{-4}$, and increasing for higher ranks $\kappa$. Thus we consider the EQS as any state minimizing $A^{(19)}_4$. Moreover, because the satisfiable criteria defining the EQS are specific to the weight of its irrep, EQS for different system sizes do not resemble each other, and so QSLs and target fidelities are incomparable between them. For this reason, we do not perform scaling analysis on the EQS, and only consider the 19-qubit case.

\subsection{Control Cost Functions}
\subsubsection{Collective control without leakage penalty}\label{sec:naive}
The simplest control strategy uses the infidelity, $C_F=1-F$, as the cost-function. While this is well suited to problems where the Hilbert space is closed under control operators, in our problem this strategy is ``na\"ive" because we perform control on a truncated Hilbert space. First-order irrep distillation, $\mathcal{H}_D^{(1)}$, is not closed under $\hat{H}(\phi,\Omega)$, and generally this will lead to leakage out of the simulated subspace. The infidelity cost function $C_F$ draws no distinction between the target and leakage subspaces of $\mathcal{H}_D^{(1)}$, treating the entire vector space as computational. We consider na\"ive control in the prediction that, by failing to penalize leakage, it will generate pulses which fail to generate the target state when implemented on expanded Hilbert spaces: $\mathcal{H}_D^{(2)}$ and $\mathcal{H}$.

\subsubsection{Collective control with leakage penalty}\label{sec:global}
\begin{figure}
    \centering
    \includegraphics[width=\linewidth]{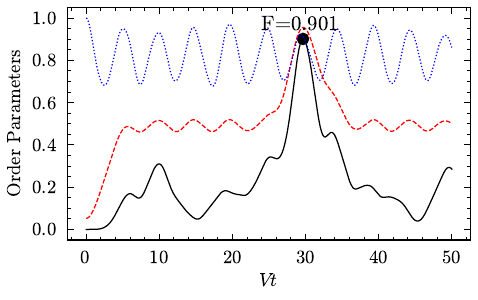}
    \caption{\label{fig:XYevo} Evolution of an initial spin coherent state $\ket{\uparrow}^{\otimes N}$ by time-independent $\hat{H}_\mathrm{int}$, modeled in $\mathcal{H}_D^{(2)}$, demonstrating approximate GHZ state formation at $T=29.65/V$ with $F=0.901$. The GHZ fidelity $F$ (solid black) peaks at the same time as the variance $\frac{4}{N^2}\Delta J_y^2$ (dashed red), as expected of a GHZ state. The population in the symmetric subspace $\langle \Pi^{(0)}\rangle$ (dotted blue) oscillates.}
\end{figure}
To ensure simulated pulses also perform well when the truncated Hilbert space is restored to full dimensionality -- that is, in the exact dynamics -- we must build into the cost function a preference against populating escape subspaces. Like na\"ive control above, we consider a ``global" control with collective Rabi rotations, generated by $\Omega(t)\hat{J}_{\phi(t)}$. We model the total Hamiltonian on $\mathcal{H}_D^{(x)}$, the $x^\text{th}$ order of irrep distillation, but its cost-function discourages performing control in the escape subspace. This cost-function is 
\begin{equation}
    C_x=1-F+P_x
\end{equation}
wherein $P_x$ approximates the fractional population-time spent in the escape subspace of $\mathcal{H}_D^{(x)}$:
\begin{equation}
    P_x\equiv \epsilon\sum_{k=1}^K \langle \Pi^{(x)}\rangle_k.
    \label{eq:P1}
\end{equation}
We normalize $\epsilon=\frac{1}{K}$ so that $P_x\in[0,1]$; GrAPE yields sufficiently many choices of pulses that convexly optimizing for $C_x$ does not reduce $F$ in the optimal pulse, and so the exact weight $\epsilon$ is unimportant (see \cref{app:Pareto}). We present results according to each truncation separately. Because IRD is a controlled approximation, we expect to be able to solve pulses in $2^\text{nd}$-order IRD which better account for leakage than does $1^\text{st}$-order. We also consider a bounded control scheme which limits $\Omega$ to the weak-driving regime, otherwise identically with ``global" control, in \cref{app:Slow}.

\subsubsection{Local control with leakage penalty}\label{sec:local}
Global control provides limited options to actively reclaim population leaking into lower irreps, since the only permutation asymmetric term in the Hamiltonian is the interaction term, $\hat{H}_\mathrm{int}$. Thus, a control scheme that actively pumps leaked populations back into the symmetric subspace should utilize local addressing, but the extensive number of control parameters involved in that level of fine-tuned control is computationally unscalable, not to mention experimentally cumbersome. But at the lowest perturbative order, IRD informs us that only a few different irreps are intrinsically coupled to the symmetric subspace, and so local control can suffice to suppress or reverse only these matrix elements. This amounts to the introduction of control terms with tensorial form $\hat{T}^{(k)}_q[\frac{N}{2};\frac{N}{2}-1,0]$ etc. In fact, the local addressing terms should \textit{only} couple the symmetric subspace to those irreps already isolated by IRD, to prevent the pump from incurring its own leakage into new irreps.

We test a method of control with simple local-addressing and few parameters by introducing a second, inhomogeneous field with a Rabi frequency $\omega$ and the same phase $\phi$ as the global field with Rabi frequency $\Omega$. This inhomogeneous field locally addresses atoms according to the net strength of their interactions with other atoms: $\bar{V}_i\equiv \sum_j V_{ij}$. Thus the local control Hamiltonian takes the form,
\begin{align}
    \hat{H}(\phi(t),\Omega(t),\omega(t))=\Omega(t)\hat{J}_{\phi(t)}+\omega(t)\hat{K}_{\phi(t)}+\hat{H}_\mathrm{int},\\
    \hat{K}_{\phi(t)}=\sum_{i=1}^N \bar{V}_i[\cos(\phi(t))\hat{\sigma}_x^{(i)}+\sin(\phi(t))\hat{\sigma}_y^{(i)}].
\end{align}
In the irrep decomposition of Hilbert space, $\hat{K}_\phi\propto \{\hat{T}^{(1)}_{-1},\hat{T}^{(1)}_{1}\}$ and we acquire matrix elements of the form
\begin{align}
    \bra{\frac{N}{2},M}\hat{K}_\phi\ket{\frac{N}{2}-x,M',u}=\\
    F_1(0;x,u)e^{i\phi}\bra{\frac{N}{2},M}\hat{T}^{(1)}_{-1}\ket{\frac{N}{2}-x,M',u}\\
    -F_1(0;x,u)e^{-i\phi}\bra{\frac{N}{2},M}\hat{T}^{(1)}_{1}\ket{\frac{N}{2}-x,M',u},
\end{align}
where $\vec{F}_1(0,1)$ is parallel to $\vec{F}_2(0,1)$ due to $\hat{K}_\phi$ being constructed from the graph $V_{ij}$. Thus, $\hat{K}_\phi$ couples significantly between the same irreps as are included in IRD, and insignificantly outside the IRD subspace. Local control using $\hat{K}_\phi$ may actively reverse leakage due to $\hat{H}_\mathrm{int}$ without causing significant leakage itself.

Other than the greater parametrization in the Hamiltonian, the local control scheme proceeds identically to the above global control, in its cost function and optimization protocol. We consider only IRD-2, since active pumping only improves the cost function if there exists an auxiliary subspace that is neither the target nor escape subspace.

\section{Results}\label{sec:res}
\begin{figure}
    \centering
    \includegraphics[width=\linewidth]{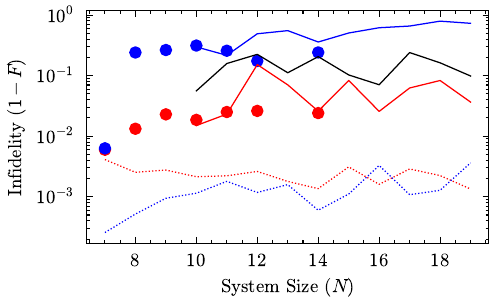}
     \caption{Preparation of GHZ states with QOC with global driving. Pulse sequences are found with GrAPE on a Hilbert space truncated according to IRD-1. The resulting infidelities are plotted as a function of the number of qubits. These same pulses are then applied to dynamics on an extended Hilbert space based on IRD-2, and for small systems sizes the exact dynamics on the full Hilbert space to test convergence. Blue: pulses designed without leakage penalty (``na\"ive control") on IRD-1 (dotted), IRD-2 (solid), and full Hilbert space (circles). Red: global pulses with leakage penalty on IRD-1 (dotted), IRD-2 (solid), and full Hilbert space for $N\le 14$ (circles), with $K=VT=N+6$. With IRD-2 we have good confidence that we are capturing the true fidelity. Black: time-independent GHZ formation in IRD-2 (compare \cref{fig:XYevo}).}
    \label{fig:NaiveF}
\end{figure}
We evaluate the quality and complexity of the control protocols as quantified by the fidelity with which they produce target states, and the quantum speed limit (QSL) to achieve that fidelity. As a benchmark, note that, given an initial spin coherent state on the Bloch sphere's equator, the OAT Hamiltonian naturally generates a GHZ state without time-dependent driving at constant time regardless of system size~\cite{Pezze18}. Given the $U(1)$ symmetry, $\hat{H}_{gOAT}$ will do the same, assuming extensive interaction magnitude; with Kac-normalized interactions as in our simulations, this becomes linear time. By initializing in $\ket{\psi(0)}=\ket{\Uparrow_x}$, evolution by $\hat{H}_{gOAT}$ acquires $\ket{\psi(T)}=\frac{1}{\sqrt{2}}(i^{N}\ket{\Uparrow_y}+\ket{\Downarrow_y})$ exactly for odd $N$. As one benchmark, for quantum control to be useful for GHZ construction on finite-range interacting systems, it must construct GHZ states with higher fidelity than passive evolution with $\hat{H}_\mathrm{int}$ can in the absence of driving. \Cref{fig:XYevo} demonstrates the time-independent dynamics generated by $\hat{H}_\mathrm{int}$ for $N=19$, evolving using IRD-2 and finding GHZ fidelity as $F\approx 0.9$ at $T\approx 30/V$. A similar simulation in IRD-1 shows $F=0.923$ around the same time. These provide performance benchmarks for QOC, which we find significantly outperforms passive evolution on the interaction graph $V_{ij}$, achieving superior $F$ over shorter $T$.

\subsection{Fidelities}\label{sec:fid}
\begin{figure}
    \centering
    \includegraphics[width=\linewidth]{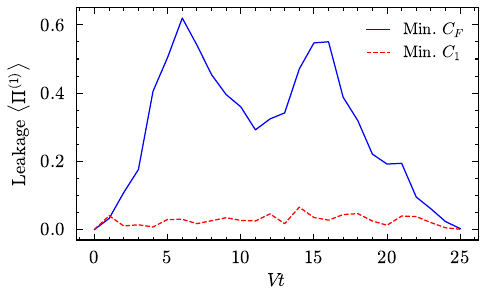}
    \caption{Population $\langle \Pi^{(1)}\rangle$ in the leakage subspace over the course of evolution. Two pulse sequences $(\vec{\phi},\vec{\Omega})$ to construct a GHZ state in $\mathcal{H}_D^{(1)}$ with $N=19$, $K=VT=25$. Solid blue pulse: ``na\"ive" control minimizing $C_F$ without a leakage penalty. Dashed red pulse: ``global" control minimizing $C_1$ with a leakage penality. The na\"ive cost function will yield artificially high fidelities, ignoring population that will quickly leak out of the target Hilbert space. }
    \label{fig:NaiveLeakage}
\end{figure}
\begin{figure*}
\centering
\includegraphics[width=\linewidth]{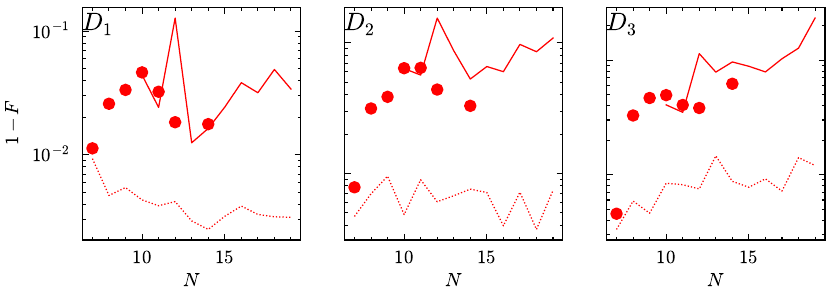}
\caption{\label{fig:DickeF} Infidelity $1-F$ in the control of Dicke states $\ket{D_k^{(N)}}$ as a function of the number of qubits, $N$. $K=N+2k+1$ pulses are employed using global control in IRD-1 with leakage penalty, for total time $VT=N+k+1$. The infidelities calculated from pulse implementation on $\mathcal{H}_D^{(1)}$ (dotted line), $\mathcal{H}_D^{(2)}$ (solid line), and full Hilbert $\mathcal{H}$ (circles) for $N\le 14$ .}
\end{figure*}
Control pulses found based on QOC in the truncated Hilbert spaces of IRD predict their own target fidelity as a component of the cost function $C$, and we test these predictions for perturbative stability by comparing QOC with higher-order IRD or to the exact dynamics on the full Hilbert space. This is done by acquiring a pulse of phases and Rabi frequencies, $\vec{v}$, solved in either IRD-1 or IRD-2. This defines the time-dependent Hamiltonian and associated the unitary evolution. To assess the accuracy of the control on the truncated Hilbert space, we apply this unitary map to an expanded Hilbert space, either by comparing IRD-1 with IRD-2 at large $N$, or the full Hilbert space $\mathcal{H}$ at small $N$. 

Firstly, as expected, na\"ive global control without a leakage penalty demonstrates untenable true fidelity. The blue data in \cref{fig:NaiveF} shows that the exact infidelity grows to $1-F>2\times 10^{-1}$ at any considerable system size $N>7$, which is not reflected in the optimistic IRD-1 cost function $C_F<5\times 10^{-3}$. This infidelity increases with $N$ furthermore; the failure of na\"ive control illustrates the necessity of preventing leakage in models of finite-range interacting systems. As \cref{fig:NaiveLeakage} elucidates, na\"ive control incurs extensive leakage because it makes no distinction between the target and escape subspaces in the modeled Hilbert space of IRD; by abusing the escape subspace for computation, na\"ive control becomes nonrobust to higher-order couplings.

On the other hand, global control with a leakage penalty using IRD-1 demonstrates good fidelity at all simulated system sizes, as shown in \cref{fig:NaiveF}, in which GrAPE pulses solved in $\mathcal{H}_D^{(1)}$ are applied at various $N$ to construct GHZ states. The IRD-1 infidelity is similarly low as in the case of na\"ive control, $1-F<5\times 10^{-3}$ at all $N$, with IRD-2 infidelity tending to be more pessimistic than the exact infidelity in the range $N\in\{10,11,12,14\}$. For this reason, we may suspect the infidelity predicted using global control with a leakage penalty for IRD-2 at $N=19$ to be too conservative. Within $N\in[7,12]$ and again at $N=14$, exact solutions of the pulses still give $F\geq 0.975$. This is equivalent to CZ fidelity $F_{CZ}=F^{1/(N-1)}=0.998$ for a 14-qubit GHZ state, better than current physical Rydberg two-qubit gates~\cite{Evered23,Tsai25}. For $N>14$ where exact diagonalization is impossible, then IRD-2 serves as a proxy check for the performance of IRD-1, as its infidelities track with those of the exact system. Comparing global na\"ive and leakage-penalty control schemes in \cref{fig:NaiveLeakage}, we find that implementing the leakage penalty $P_1$ is effective and necessary in guaranteeing real fidelity when truncating to $\mathcal{H}_D^{(1)}$.

\begin{figure}
    \centering
    \includegraphics[width=\linewidth]{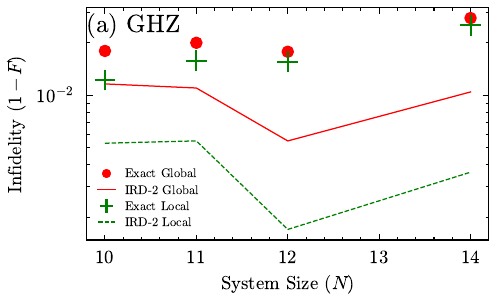}
    \includegraphics[width=\linewidth]{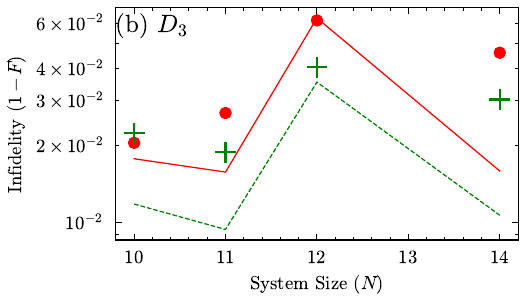}
    \caption{Target state infidelity $1-F$ from control pulses calculated on IRD-2, across system sizes $N$. Top: $\ket{\Psi_\mathrm{tar}}=\ket{\mathrm{GHZ}}$; bottom: $\ket{\Psi_\mathrm{tar}}=\ket{D_3}$. Two control strategies: global with leakage penalty (red), and local (green). Infidelity calculated from implementing pulse on truncated $\mathcal{H}_D^{(2)}$ (solid lines) or full $\mathcal{H}$ (circles).}
    \label{fig:GHZ2}
\end{figure}

Our results for preparing Dicke states corroborate previous QOC results~\cite{Muratori25}. The states $\ket{D_1}$ ($W$-states) remain controllable at large $N$, but lower Dicke states $\ket{D_k}$ with $k>1$ exhibit entanglement structures which grow far quicker with system size and result in greater sensitivity to finite-range interactions. The ease of constructing $\ket{D_1}$, and comparative difficulty of constructing $\ket{D_3}$, becomes apparent in \cref{fig:DickeF}, wherein exact infidelities at $N=14$ are $F\approx 0.98$, $\approx0.96$, $\approx 0.94$ for $\ket{D_1}$, $\ket{D_2}$, and $\ket{D_3}$ respectively. Moreover while the first Dicke state shows an overall flat or downward trend in infidelity as system size grows (at all levels of approximation), the third Dicke state shows growing infidelity with system size. This implies that above a certain threshold in $N$, the interaction graph $V_{ij}$ will become insufficiently well-connected for symmetric control. Nevertheless, the exact fidelities do not underperform relative to the predicted performance of deterministic physical circuits. According to a measurement-free algorithm~\cite{Bartschi22}, preparation requires $\mathcal{O}(Nk)$ two-qubit gates, which we treat as exactly $Nk$ for simplicity. With current two-qubit gate fidelity $F_{CZ}=0.9971$ from \cite{Tsai25}, then $(F_{CZ})^{3N}=0.885$ for $N=14$, much worse than the exact infidelities of $\ket{D^{(14)}_3}$ in \cref{fig:DickeF,fig:GHZ2}. However, measurement-free protocols are not necessarily the optimal method for circuit-construction of Dicke states, as \cite{Yu26} points out.

Extending global control with leakage penalty to IRD-2 provides minor benefit to actual infidelities, shown by \cref{fig:GHZ2}, in which the exact infidelities of IRD-2 global pulses in the range $N\in\{10,11,12,14\}$ are only slightly lower than the corresponding infidelities from IRD-1 pulses in \cref{fig:NaiveF}. Moreover, the use of local control only marginally improves actual fidelity over global control, though when infidelities are calculated within IRD-2 the local control appears to significantly improve upon the global control. For more difficult states like $\ket{D^N_3}$ however, even minor improvements from higher-order IRD and local control prove invaluable: pulses with equivalent $(K,T)$ at $N=14$ give exact infidelity $1-F\approx 6\times 10^{-2}$ on IRD-1 global control, but $1-F\approx 3\times 10^{-2}$ on IRD-2 local control, a significant improvement shown in \cref{fig:GHZ2}. The $D_3$ fidelity from local control is thus highly competitive relative to expected experimental fidelity on gate-based hardware. This suggests that investing in experimentally demanding individual-addressing mechanisms to negate dynamical asymmetries, without sophisticated protocols for individual atom control, will only merit the effort when pursuing highly detailed, resourceful states like large-$k$ Dicke states.

For EQS, the large leakage out of the Dicke manifold during control, as indicated by QSL investigations on the EQS target, signals severe, irretrievable population loss on the extended Hilbert space. We find $C_1$ for the EQS becomes insensitive to changes in timescale for $VT\gtrsim 45$, and to changes in pulse number for $K\gtrsim 100$, but $C_1$ does not approach zero beyond these thresholds, as we demonstrate in \cref{app:EQS}. Pulses solved in IRD-1, when applied to higher-order IRD, demonstrate universally low fidelity. Robust and efficient protocols for arbitrary EQS construction have recently been developed using a trotterization of $\hat{H}_0(t)$~\cite{Denis26} which our own symmetric numerics corroborate. This proves the extreme case of the principle already shown by $\ket{D_3}$, that global control with dipolar interactions in 2D cannot reach full controllability on the symmetric subspace with efficiently solvable pulses, although it can well-approximate many desirable states and herald its own instability via $P_1$ and $P_2$. This failure of QOC on the EQS aligns with prior results~\cite{Carrera26}, that globally addressed control on Rydberg arrays interacting with $1/r^6$ van der Waals interactions gives infidelity which scales exponentially with the symmetric target's entanglement entropy. Since extremal quantum states generically have near-maximum entanglement for a symmetric state~\cite{Goldberg20}, the findings of \cite{Carrera26} would suggest that constructing an EQS is impossible on this family of control in the limit of large $N$.

\subsection{Quantum Speed Limits}\label{sec:QSL}
\begin{figure}
    \centering
    \includegraphics[width=\linewidth]{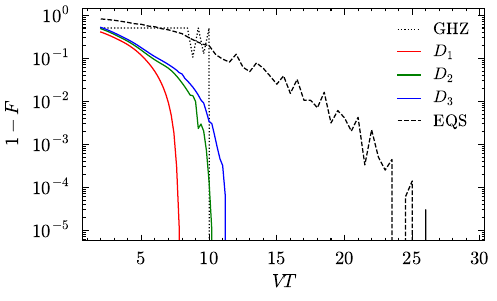}
    \caption{The Quantum Speed Limit of EQS, Dicke, and GHZ states controlled by an all-to-all symmetric Hamiltonian with $ZZ$ OAT interactions, for $N=19$, $K=20$. Symmetric control can construct arbitrary states in $\mathcal{H}_{N/2}$ in $VT\lesssim N$ with arbitrary fidelity $F$.}
    \label{fig:SymQSL}
\end{figure}

For a finite dimensional Hilbert space of dimension $d$, a pure state is specified by $2d-2$ real parameters, which is the minimum number of control parameters in the waveform $\vec{v}$ needed to specify $|\Psi_\mathrm{tar}\rangle$. In practice, the control landscape is more robust with a few additional parameters, e.g., $||\mathcal{V}||=2d$. For symmetric control with parameters $(\phi,\Omega)$, this equates to $K=N+1$. By the same token, with Kac normalization it is expected that symmetric dynamics achieve full control for $T\propto N/V$~\cite{Chaudhury07}. We demonstrate these speed limits on a symmetric system controlled by $\hat{H}_0(t)$ in \cref{fig:SymQSL}, showing vanishing infidelity on all target states even at $K=N+1$. For any choice of evolution time beyond a state's QSL, symmetric GrAPE can solve a pulse which gives arbitrarily low infidelity.

\begin{figure*}
\centering
\includegraphics[width=\linewidth]{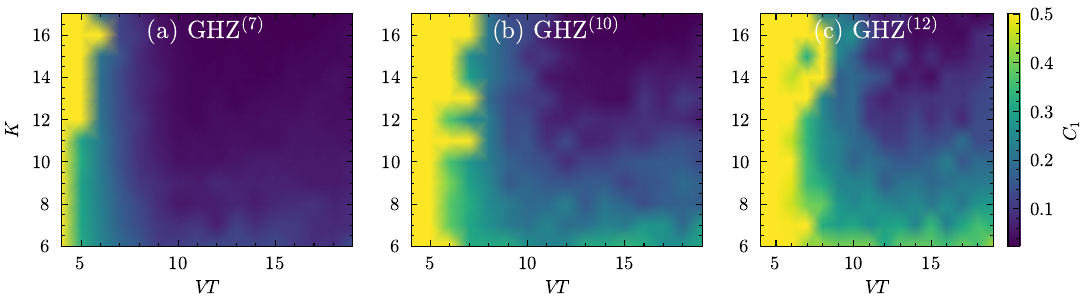}
\caption{\label{fig:GHZMT} Global-control cost function $C_1$ to construct $\ket{\mathrm{GHZ}}$ for ranges $K\in[6,17]$, $VT\in[4,19]$, with $N\in\{7,10,12\}$. Minimum $(K,T)$ to saturate $C_1\approx0$ for a given $N$ represents the QSL, which for the $N$ values shown above, are $(K,VT)\gtrsim N+6$.}
\end{figure*}

With finite-range interactions, the QSL is complicated by the fact that the target states are confined to a subspace of the full Hilbert space, and the complexity of the control will depend on how much of the additional Hilbert space we include in the optimization. As $P_1$ and $P_2$ are not guaranteed to approach zero with growing $(K,T)$, then for dipolar coupling the QSL may not converge, defined as the $(K,T)$-value beyond which one may find $C<\epsilon$ for arbitrary small $\epsilon$. We instead decouple the QSL from the final fidelity by considering it merely as the point in $(K,T)$ beyond which $C$ plateaus. In that sense, we find accounting for leakage drastically lengthens the QSL in both time $T$ and complexity $K$. The three unconstrained control schemes -- na\"ive-global control without leakage penalty, global control with leakage penalty, and local control -- do not meaningfully differ in their QSLs for GHZ construction. All three protocols reach plateaus in their cost functions for $(K,VT)\gtrsim N+6$, which comports with previous results that GHZ-states require linear time and complexity from global control under Kac normalization. \Cref{fig:GHZMT} shows the linear scaling, including the fact that much of the added time and complexity of control with the resonant dipole interaction arises from combating leakage.
\begin{figure*}
\centering
\includegraphics[width=\linewidth]{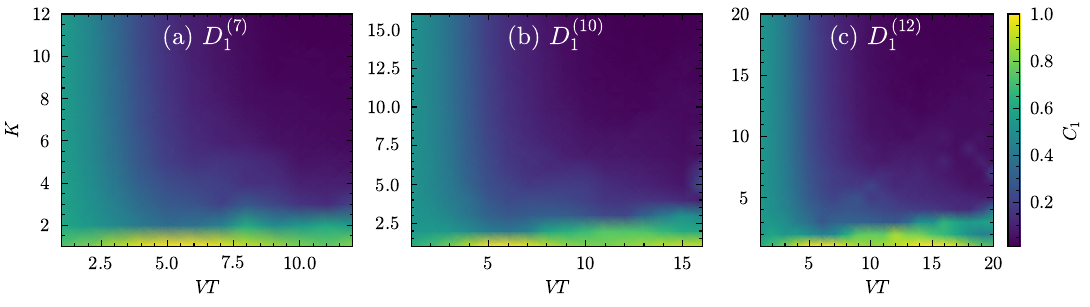}
\caption{\label{fig:DickeMT} Global-control cost function $C_1$ to construct $\ket{D_1^{(N)}}$ for ranges in $(K,T)$, with $N\in\{7,10,12\}$. Minimum $(K,T)$ to saturate $C_1\approx0$ for a given $N$ represents the QSL, which for the $N$ values shown above, are approximately $K\gtrsim N+3$, $VT\gtrsim N+2$.}
\end{figure*}

\begin{figure*}
\centering
\includegraphics[width=\linewidth]{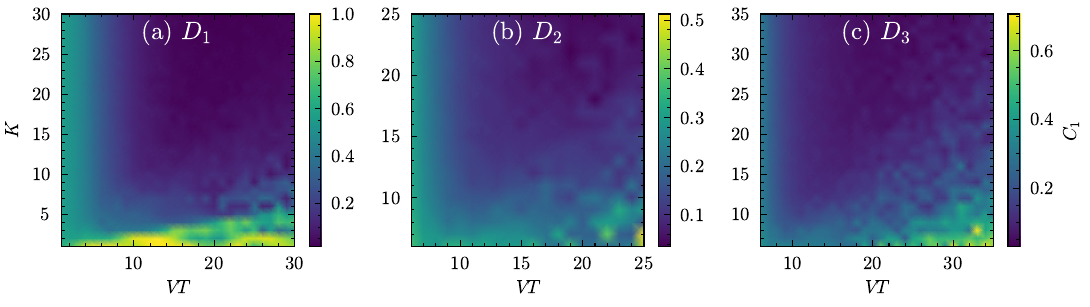}
\caption{\label{fig:DickeMT2} Global-control cost function $C_1$ to construct $\ket{D_k^{(19)}}$ for ranges in $(K,T)$. Minimum $(K,T)$ to reach plateau in $C_1$ represents the QSL, showing apparent linear growth of QSL with $N,k$. Convergence for $K\gtrsim N+2k+1$, $VT\gtrsim N+k+1$.}
\end{figure*}

The approximate QMBS and spectrum of $\hat{H}_\mathrm{int}$ change only perturbatively as the IRD order increases, thus moving from $\mathcal{H}_D^{(1)}$ to $\mathcal{H}_D^{(2)}$ negligibly impacts the QSL~\cite{Wiedmann26}. But the target state heavily influences the QSL, as shown in the the case of symmetric all-to-all interactions, \cref{fig:SymQSL}. The QSL for the $W$-state is evidently linear with $N$ as shown in \cref{fig:DickeMT}, and demonstrates a well-optimized $C_1$ under global control for $K\gtrsim N+3$, $VT\gtrsim N+2$. The QSL for Dicke states grows slightly with the Dicke lowering $k$, but in all cases the plateau region in $(K,T)$ is approximately equal to that of the GHZ, in keeping with the symmetric case. However, as \cref{fig:DickeMT2} shows, the optimum $C_1$ accessed by global control is suppressed with increasing $k$, suggesting increasingly unavoidable leakage in the course of evolution.

\section{Physical Implementation}\label{sec:disc}
\begin{table}
    \centering
    \begin{tabular}{|c c|c c c|c c}
         \hline
         Species & Embed & $\tau$($\mu$s) & $a$($\mu$m) & $V_{1,2}$(MHz) & $\eta$ & $\exp(-1/\eta)$\\
         \hline
         %$^{87}$Rb\cite{Bornet23} & $\substack{60S_{1/2}\\ 60P_{3/2}}$ & 230 & 15 & $\pi$ & 2.3 & 0.65\\
         $^{87}$Rb\cite{Chen26} & $\substack{60S_{1/2}\\ 60P_{1/2}}$ & 230 & 10.8 & $7.54$ & 11 & 0.91\\
         $^{133}$Cs\cite{Zhang24} & $\substack{53D_{3/2}\\ 54P_{1/2}}$ & 55 & 8 & 113 & 19.8 & 0.951\\
         $^{133}$Cs\cite{Zhang24} & $\substack{63P_{3/2}\\ 64S_{1/2}}$ & 112 & 16 & 9.916 & 3.51 & 0.752
    \end{tabular}
    \caption{The speed figure of merit $\eta$ for various experimental and hypothetical Rydberg platforms. The species and embedded qubit manifold determine the single-qubit lifetime $\tau$, and with specific interatomic spacing $r$, then the interaction frequency $\mathcal{J}$ becomes determined.}
    \label{tab:speeds}
\end{table}
We consider physical implementation based on dipolar interactions between atoms. As our qubit we consider two Rydberg states of opposite parity, $\ket{\uparrow}=\ket{n,l}$, $\ket{\downarrow}=\ket{n',l\pm 1}$, with a lattice spacing $a$, giving a dipolar interaction strength between adjacent atoms $V_{1,2}=d^2/a^3$, where $d$ is the dipole matrix element. In experiment, it is important to choose $a$ to maximize $V_{1,2}$ without bringing atoms close enough that second-order dipole-dipole (Van der Waals) interactions become non-negligible. Examples based on recent experiments are shown for Rb~\cite{Chen26} and Cs~\cite{Zhang24} in \cref{tab:speeds}.

Fundamentally, even for perfect control, the fidelity will be limited by the finite lifetime of the Rydberg states. For a given atomic species and orbital manifold for the embedded qubit, the single-qubit lifetime $\tau$ is the lesser of the two states' lifetimes as we calculate with ARC~\cite{ARC}; the collective lifetime $\tau_\mathrm{col}\equiv \frac{\tau}{N}$ follows from an exponential decay model of Rydberg state leakage. We compare our simulated QSL for GHZ-construction, $VT=25$, against $\tau_\mathrm{col}$ by equating $V_{1,2}=\frac{d^2}{a^3}=\frac{V}{\mathcal{N}}\approx 0.666V$ to experimental frequencies. We define a speed figure of merit,

\begin{equation}
    \eta\equiv \frac{V\tau_\mathrm{col}}{V_{1,2}T}=\frac{\mathcal{N}\tau}{NT},
\end{equation}
representing the number of consecutive control pulses we can perform on the Rydberg ensemble before any of the atoms are likely to leak out of their qubit manifolds. 

The unbounded control strategies laid out above generate pulses with large Rabi frequencies $\Omega\gg V$, sometimes greater by a factor of 10. Microwave driving can achieve these Rabi frequencies for the large electric dipole moments associated with Rydberg atoms. While we have considered piecewise constant (discontinuous) control in our numerics, in practice this is not required. Smooth pulse design is also possible with other algorithms, e.g., GOAT~\cite{PhysRevLett.120.150401}. Setting aside these limitations, we show the figure of merit for a selection of Rydberg platforms in \cref{tab:speeds}.

If the dominant path of Rydberg decay is leakage out of the qubit manifold, as opposed to a bit-flip within the qubit, then we may model this decay as a simple suppression of the target fidelity by a factor $e^{-1/\eta}$. Indeed, the bit-flip rate, considered as the transition rate $\ket{\uparrow}\to\ket{\downarrow}$ or vice versa~\cite{ARC}, in any qubit manifold shown above, at either near-zero temperature or room-temperature, is always significantly slower than the limiting lifetime. The worst case is at room-temperature, the transition $^\mathrm{Cs}:64S_{1/2}\to 63P_{3/2}$ occurs at a rate $7.73$ times slower than $\tau_\mathrm{col}^{-1}$. This implies that amplitude damping noise is much slower than leakage, and thus negligible during state-construction. This is also the embedding with the worst $\eta$ in \cref{tab:speeds}, and so all the cryogenic Rydberg qubits, and the better-performing Rydberg qubits generally, also have clean decay characteristics and minor fidelity-suppression.

\section{Conclusion}\label{sec:conc}
In this investigation we studied the use of quantum optimal control to prepare quantum states in the Dicke manifold based a 2D-lattice of Rydberg atom qubits interacting via an $XX$-Hamiltonian with finite-range (dipolar) and driven Rabi oscillations. Exploiting finite-size quantum many-body scars (QMBS) and irreducible representation distillation (IRD), we perturbatively expressed the long-range interactions as a Hamiltonian-driven leakage out of the symmetric subspace. By integrating this error model into our strategy of gradient ascent pulse engineering (GrAPE), we have solved leakage-aware control protocols which can construct several resourceful quantum states with competitive fidelity without resorting to high-resolution local control or many-parameter individual addressing schemes. 

We achieve high fidelity for GHZ states and W-states with $N=19$ qubits, with only global control, competitive with what can be achieved with quantum circuit-based protocol. Cost functions which penalize leakage out of the symmetric subspace predict their own success or failure without querying the full Hilbert space, a highly useful feature for efficient simulation and control. Increasing quantum complexity requires longer and more complex control waveforms. Highly quantum states in the symmetric subspace, such as the extremal quantum state (EQS), remain intractable, suggesting that global control with finite-range interactions cannot perfectly span the symmetric subspace. This may be explained by a group-theoretic description of the time-evolution generated by the full Hamiltonian, $\hat{H}(\phi(t),\Omega(t))$, which differs from that by $\hat{H}_0(t)$ in its ineliminable couplings between irreps of different weights. Since commutation relations between operators coupling multiple irreps of $SU(2)$ are far more intricate than between operators acting on a single such irrep, the path to proving the (non-)compactness of the set of unitaries generated by $\hat{H}(\phi(t),\Omega(t))$ would be arduous. However, we see suggestions for the non-simple topology of this set, in the fact that our control pulses consistently give $1-F>0$ even when solved and then tested on the same Hilbert space. \Cref{fig:SymQSL} demonstrates that exact solutions $1-F=0$ are possible when the group conserves $J$, but ``na\"ive" control does not give this result even on IRD-1 in \cref{fig:NaiveF}. Thus it may be that certain states within an irrep, and certain superpositions between irreps, are simply unreachable without fine-grained locally addressed control.

We have demonstrated the complicating effect of permutation asymmetry on the quantum speed limit (QSL) for control, which suppresses the maximum fidelity and increases the dependence of the simulated cost function on pulse initialization. Moreover, because the QSLs for many states entail more complex pulse sequences and longer control times than they do on the corresponding collective Hamiltonian $\hat{H}_{gOAT}$, it follows that pulses solved by simulating a collective system will not give good experimental fidelities on the finite-range system. For this reason, it is necessary when solving pulses to model the finite interaction range, and furthermore, to penalizing leakage from the symmetric subspace. Despite the increase in the QSLs, the resonant dipole-dipole graph retains sufficient connectedness within our range of investigation, that $VT\propto N$, which is the same fundamental QSL scaling as for collective spin systems. As for the leakage itself, our fidelity results show that it is possible to prevent algorithmically without extensively parametrized Hamiltonians (i.e., without fine-grained individual addressing), and thus to construct several resourceful states with high fidelity. These solutions are enabled by IRD, which exploits the irrep decomposition of Hilbert space and perturbation theory to treat the finite-range interactions as generators of leakage errors out of the symmetric subspace. This leakage model of Hamiltonian dynamics is justified in turn by the chaos of the Hamiltonian, implying that leaked population will only return to the symmetric subspace after exponential time.

More complex control strategies may enable an extension of the range of states and gates that can be performed on the Dicke manifold. We found that local control only leads to minor improvement over global control protocols. There remain open questions on how best to penalize leakage or reclaim leaked population. Finally, this theory has only explored open-loop control, without measurement or feedback. As the role of measurement-based feedback in quantum control problems becomes better appreciated and understood, new space opens up to implement those strategies on the resonant dipole-dipole Rydberg array.

Finally, it bears repeating that the dynamics simulated here, characterized by power-law decaying interactions on a lattice with $\alpha=3$ and $D=2$, lie outside the strong long-range regime~\cite{Lerose19,LP20,Mori19} and therefore benefit only weakly from protections of quantum many-body scars (QMBS)~\cite{LP25} and energy gaps~\cite{Kuriyattil25}. As neutral atomic platforms advance and 3-dimensional lattices scale up, experiments may soon reach the long-range threshold of $\alpha=D$, where IRD will acquire a scalable remit and the control protocols laid out in this investigation will provide even better fidelities.

\section{Acknowledgments}
We thank Andrew K. Forbes for helpful discussions. This material is based upon work supported by the National Science Foundation, grant PHY-2210013 and the Air Force Office of Scientific Research, under Award Numbers FA9550-22-1-0498 and FA9550-20-1-0123. PMP acknowledges support from the Royal Society through a University Research Fellowship (Grant No. URF\textbackslash R1\textbackslash252030).

\bibliography{main}

\appendix
\section{Laplacian Gap Decay}\label{app:Lap}
\begin{figure}
    \centering
    \includegraphics[width=1\linewidth]{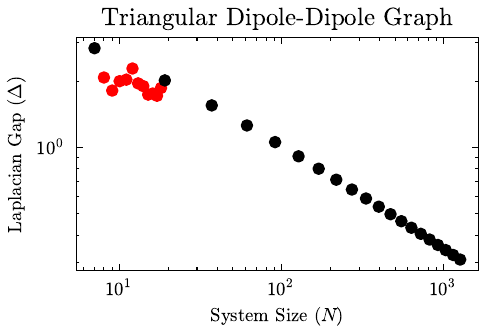}
    \caption{Laplacian gap $\Delta$ for the triangular lattice, dipole-dipole interaction graph $V_{ij}$, as a function of system size $N$. Black points are complete hexagons with $N=3L^2+3L+1$, red points are incomplete hexagons for $N\in\{8,...18\}$. Logarithmic axes illustrate the asymptotic decay of $\Delta$ in the thermodynamic limit.}
    \label{fig:LapGap}
\end{figure}
The Laplacian matrix describes the connectedness of the interaction graph,
\begin{align}
    \Lambda_{ij}=\bar{V}_i\delta_{ij}-V_{ij},\\
    \bar{V}_i=\sum_j V_{ij},
\end{align}
and is positive semidefinite: $\Lambda\geq 0$, furthermore its lowest eigenvalue is always $\lambda_0=0$. Consequently, the spectral gap between the Laplacian's ground state and first excited state is $\Delta\equiv \lambda_1-\lambda_0=\lambda_1$. As stated in the main text, $\hat{H}_\mathrm{int}\approx \hat{H}_{gOAT}$ for large $\Delta$. However, in the large-system limit of increasing $N$, $\Delta$ only remains large if $\alpha<D$ for interaction decay exponent $\alpha$ and lattice dimension $D$; dipole-dipole interactions are defined by $\alpha=3$, and a triangular lattice has $D=2$, meaning $V_{ij}$ falls outside the strong long-range interaction regime~\cite{Defenu24,LP20}, and $\lim_{N\to\infty}\Delta=0$, which the log-log scaling in \cref{fig:LapGap} demonstrates~\cite{Kuriyattil25}.

\section{Dual Cost Optimization}\label{app:Pareto}
\begin{figure}
    \centering
    \includegraphics[width=\linewidth]{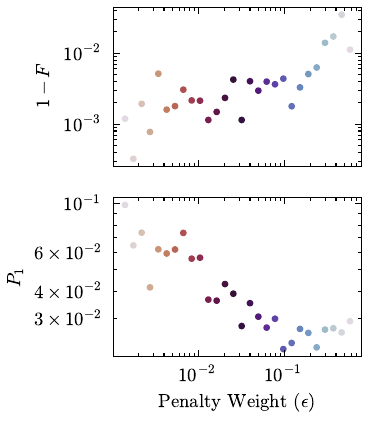}
    \caption{The Pareto front of dual optimization in $1-F$ and (unweighted) $P_1$ as a function of the penalty weight $\epsilon$. Global control targeting a GHZ state with $N=19$, $K=T=25$.}
    \label{fig:Pareto}
\end{figure}
As the cost functions $C_1$ and $C_2$ entail dual-cost optimization, it is important to appropriately weigh the two components to ensure GrAPE does not sacrifice $F$ to minimize $P_x$ nor vice-versa. Here we perform a Pareto front analysis to find $\epsilon$ that jointly minimizes both $1-F$ and $P_1/\epsilon$, in the expectation that the optimum depends on $N$, $K$, and possibly $\ket{\Psi_\mathrm{tar}}$. \Cref{fig:Pareto} shows an example of such a Pareto front, which finds the ``elbow" of the front around $\epsilon\approx 0.1$ for $N=19$, $K=T=25$, and $\ket{\Psi_\mathrm{tar}}=\ket{\mathrm{GHZ}}$. Sampling noise in the initial sequence $\vec{v}_0$ dominates the variation in $1-F$ and $P_1$, such that the specific value of $\epsilon$ affects the cost function less than the optimization trajectory in $\mathcal{V}$. Thus, we lost insignificant utility by simulating control with $\epsilon=\frac{1}{K}$, knowing it is close to optimal.

\section{Extremal Quantum State}\label{app:EQS}
\begin{figure}
    \centering
    \includegraphics[width=\linewidth]{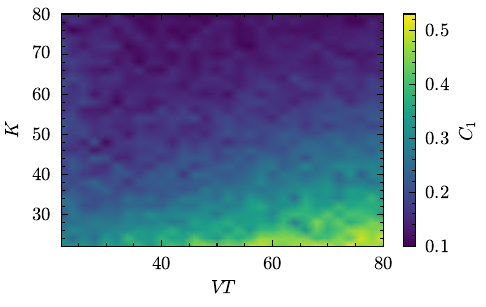}
    \includegraphics[width=\linewidth]{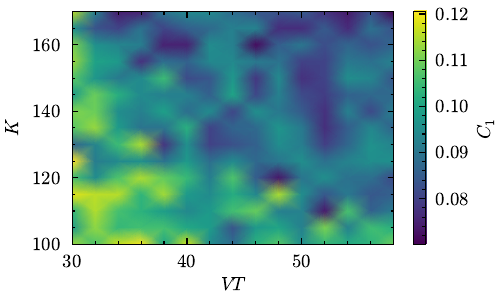}
    \caption{Global control cost function $C_1$ to construct $\ket{\mathrm{EQS}}$, across two ranges in $(K,T)$. The best cost function achieved remains bounded by $C_1\geq 0.07$, suggesting extreme leakage under evolution.}
    \label{fig:EQSMT}
\end{figure}

The EQS shows a severe suppression of $C_1$ across a wide range of $(K,T)$, reaching its optimum around $VT\approx[45,50]$, but only slowly, asymptotically approaching $C_1\to0.07$ at very large $K>100$, all shown in \cref{fig:EQSMT}. Such complex pulses are computationally expensive to solve even under IRD. The long times and potentially high bandwidths needed for implementing the required time-dependent waveforms may limit experimental feasibility. Even in theory, global control cannot avoid leakage from the symmetric subspace in the course of constructing extremal quantum states.
\begin{figure}
    \centering
    \includegraphics[width=\linewidth]{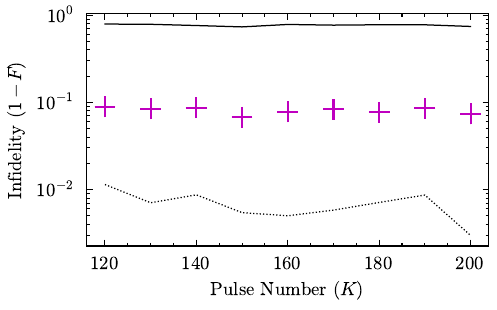}
    \caption{Infidelity $1-F$ of EQS, across a range of pulse-numbers $K\in[120,200]$ for $N=19$, $VT=48$. Pulses from global control on IRD-1 (dotted), tested against evolution in IRD-2 (solid). Magenta crosses: $C_1$ of global control pulse.}
    \label{fig:EQSF}
\end{figure}

Beyond the threshold of $(K,T)$ described above, IRD-1 shows $1-F\lesssim 10^{-2}$ consistently, but gives no evidence of infidelity approaching 0. When these intricate high-$K$ pulses are tested on IRD-2 however, the infidelity becomes untenably large: $1-F\approx 1$, shown in \cref{fig:EQSF}. The high cost function indicates that the pulse is unstable to perturbative order, and increasing the order of perturbation theory confirms this nonconvergence.

\section{Slow Control}\label{app:Slow}
We considered an alternative control philosophy, akin to adiabatic control, to preserve the QMBS by limiting the strength of $M$-level couplings via $\hat{J}_\phi$. We attempted this by limiting the magnitude of $\Omega$ to below a threshold of perturbative couplings. In the presence of unbounded $\Omega$, the QMBS becomes locally broken within specific energy regions by excited-state quantum phase transitions (ESQPT)~\cite{LP25}. If a control pulse exploits levels in such regions, this precipitates leakage out of the QMBS, and consequently the symmetric subspace. The ESQPT emerges only for nonperturbatively large $\Omega$, and so we determined the nonperturbative threshold by solving the eigenspectrum $\ket{n}\in \mathcal{H}_D$ such that $\hat{H}_\mathrm{int}\ket{n}=E_n\ket{n}$ and $\hat{J}_z\ket{n}=M_n\ket{n}$, then calculating couplings out of the QMBS via $\hat{J}_\phi$,
\begin{equation}
    \nu_\mathrm{max}=\max_{\substack{n\in \mathrm{QMBS},\\ m\not\in \mathrm{QMBS}}} \left|\frac{\bra{m}\hat{J}_\phi\ket{n}}{E_n-E_m}\right|.
\end{equation}
Because $\hat{J}_\phi\propto \{\hat{J}_x,\hat{J}_y\}$, then $\bra{m}\hat{J}_\phi\ket{n}\neq 0$ only if $M_m=M_n\pm 1$. Thus, we constrain $\Omega$ such that $\Omega \nu_\mathrm{max}<1/2$, so that $\Omega\hat{J}_\phi$ acts as a weak time-dependent perturbation on the static QMBS. This constraint differs from adiabatic control chiefly in the use of piece-wise constant Hamiltonian pulses, which entail differential discontinuities in $\hat{H}(t)$ as a function of time; adiabatic control demands that $\frac{d}{dt}\hat{H}(\phi(t),\Omega(t))$ be very small, but we place no constraints on $|\phi_{k+1}-\phi_k|$. Nevertheless, in terms of discrete derivatives, we satisfy a looser slow-change condition, since $||\hat{H}_{k+1}-\hat{H}_k||$ is limited by $\max(\Omega_{k+1},\Omega_k)$.

Because constraining the control to the weak-driving regime extends the quantum speed limit (QSL), then to keep control well below the lifetime of Rydberg states (as do the authors of \cite{Muratori25} on a short-range Rydberg ring-lattice), we select an initial state with high mutability: $\ket{\psi_0}=\ket{\Uparrow_x}=\ket{\uparrow_x}^{\otimes N}$ rather than $\ket{\Uparrow_z}$ as for other protocols. Previous research~\cite{MunozArias23} has shown that $\ket{\Uparrow_x}$ has a high QSL for generating spin-squeezing and GHZ states on symmetric Hamiltonians, to which our control Hamiltonian is closely related (see \cref{fig:XYevo}). This speed of controllability is attributed to the placement of $\ket{\Uparrow_x}$ at the saddle-point of symmetric Hamiltonians' energy phase-space, granting it a diverging Jacobian. In fact, such an initialization may generate GHZ states with a trivial GrAPE sequence of a single pulse parametrized by the dynamical quantum phase transition (DQPT)~\cite{MunozArias23}, for which reason we refrain from initializing thus with other control schemes.

It bears mentioning that $\nu_\mathrm{max}$ varies with IRD order, and the particular eigenstates $\ket{n},\ket{m}$ for which $\nu_\mathrm{max}=\bra{m}\hat{J}_\phi\ket{n}$ are unstable with increasing order. When $\hat{H}_\mathrm{int}$ is truncated to $\mathcal{H}_D^{(1)}$, then the QMBS states most strongly coupled are those with $M=\pm\frac{19}{2}$ and $E\approx 0$, whereas in $\mathcal{H}_D^{(2)}$, the strongest coupling shifts to the QMBS states with $M=\pm\frac{17}{2}$ and $E\approx-1$. This is due to level-splitting between degenerate irreps of the same spin-number, and implies that control schemes which isolate a subspace of the QMBS for computation by penalizing strongly coupled eigenspaces within the QMBS are unstable to perturbation theory, and thus nonscalable.

\begin{figure}
    \centering
    \includegraphics[width=\linewidth]{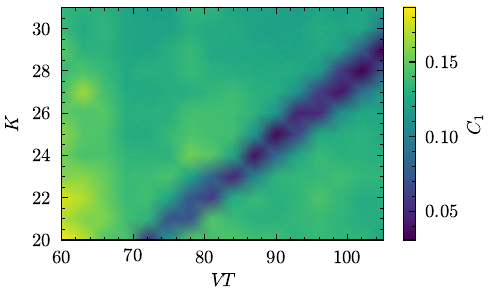}
    \caption{Slow control cost function $C_1$ to construct $\ket{\mathrm{GHZ}}$ for ranges in $(K,T)$, for $N=19$. Cost function minimized around $VT\approx 3.6K$, suggesting QSL dependent on resonance between timestep $\delta t$ and maximum driving $\nu_\mathrm{max}$.}
    \label{fig:SlowQSL}
\end{figure}
Simulations of slow control reveal a highly protracted QSL: to construct an $X$-aligned GHZ-state $\frac{1}{\sqrt{2}}(\ket{\Uparrow_x}+\ket{\Downarrow_x})$ from an $X$-polarized coherent state on $N=19$, we find an optimum time of evolution $VT\approx 3.6K$, shown in \cref{fig:SlowQSL}. This resonance relation demands that each unitary pulse evolve over $\delta t\approx 3.6/V$, which indicates fictitious fidelity due to a failure of the approximation over long time-steps. Even this evolution time benefits from the high transport speed along the Bloch-spherical equator, because attempts to target or initialize $Z$-aligned states fail to reach the QSL for all $K\leq 50$, $T\leq 550$. We explain this difficulty of control by the relative magnitudes of the two terms in $\hat{H}(\phi,\Omega)$: both $\ket{\Uparrow_z}$ and $\ket{\Downarrow_z}$ are at an extreme energy in $\hat{H}_\mathrm{int}$, but distantly separated from other states in the computation space, and due to quantum-number constraints, $\hat{J}_\phi$ cannot transport between these states directly, requiring instead an indirect route looping across the whole spectrum. Because of the low density of states near $\ket{\Uparrow_z}$ and $\ket{\Downarrow_z}$, level-crossings in that spectral band are rare, and require large $\Omega$ to induce. Thus slow control's QSL is especially sensitive to $\langle J_z^2\rangle$ in either the initial state or the target, with highly $Z$-polar sectors taking longer to control.

\begin{figure}
    \centering
    \includegraphics[width=\linewidth]{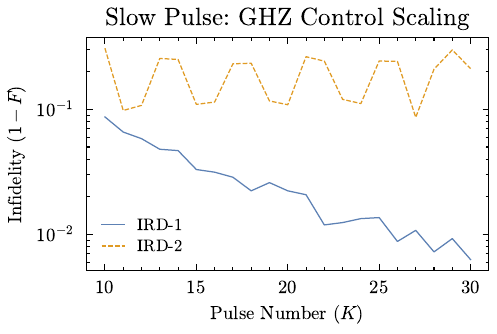}
    \caption{Infidelity $1-F$ of $X$-GHZ, across a range of pulse-numbers $K\in[10,30]$ for $N=19$, $T=3.6K$. Target infidelities taken from global control on IRD-1, tested against evolution in IRD-2.}
    \label{fig:SlowF}
\end{figure}
As for state fidelity, slow control proves non-robust to approximation order, as shown in \cref{fig:SlowF}, where pulses solved in IRD-1 show improving fidelity $F$ with pulse number $K$, but this improvement does not carry to the extended Hilbert space of IRD-2, which shows $F\in[0.7,0.9]$ consistently regardless of pulse complexity or time. This indicates that by setting $\nu_\mathrm{max}$ according to spectrum-sensitive couplings, slow control bumps against the boundaries of $\mathcal{H}_D^{(1)}$ in order to return to the symmetric subspace, but these boundaries are fictitious due to finite perturbative order. Secondly, the long timesteps $\delta t=3.6/V$ reduce the usefulness of the leakage penalty defined in \cref{eq:P1}; when $\delta t\lesssim 1$, then $P_1$ forms a Riemann-approximation of a time-integral:
\begin{equation}
    P_x=\frac{1}{K}\sum_{k=1}^K \langle \Pi^{(x)}\rangle_k\approx \frac{1}{T}\int_0^T\bra{\psi(t)} \Pi^{(x)}\ket{\psi(t)}dt,
\end{equation}
but this approximation fails when $VT\gg K$, because $\langle \Pi^{(x)}\rangle(t)$ may oscillate in the middle of a pulse without that oscillation registering in the finite sum of $P_x$. A more precise leakage penalty may be calculated by sampling multiple times across each pulse in the sequence, improving the Riemann integral at the cost of simulation complexity. Altogether, slow control shows no compelling advantage over unbounded control schemes, prior to experimental limitations.
\end{document}